\documentclass[12pt]{JHEP3}
\usepackage{graphicx}
\usepackage{epsfig}
\usepackage{amssymb}
\usepackage{bm}
\usepackage{amsmath}
\newcommand{\bea}{\begin{eqnarray}}
\newcommand{\eea}{\end{eqnarray}}

\newcommand{\pa}{\partial}
\newcommand{\bra}{\langle}
\newcommand{\ket}{\rangle}
\newcommand{\Slash}[1]{\ooalign{\hfil /\hfil\crcr$#1$}}

\makeatletter
\@addtoreset{equation}{section}
\makeatother

\title{
Vacuum Instability in Electric Fields
via AdS/CFT: Euler-Heisenberg Lagrangian and 
Planckian Thermalization
}
\author{
Koji Hashimoto$^{1}$\footnote{Also at {\it Mathematical Physics Lab., RIKEN Nishina Center,
Saitama 351-0198, Japan}}
and Takashi Oka$^{2}$
\\

$^1$
{\it Department of Physics, Osaka University,
Toyonaka, Osaka 560-0043, Japan}\\
E-mail: \email{koji(at)phys.sci.osaka-u.ac.jp}\\ 
$^2$
{\it Department of Applied Physics, University of Tokyo, 
Tokyo 113-8656, Japan}\\
E-mail: \email{oka(at)ap.t.u-tokyo.ac.jp}\\
}

\abstract{
We analyze vacuum instability 
of strongly coupled gauge theories
in a constant electric field using AdS/CFT correspondence. 
The model is the ${\cal N}=2$ 1-flavor supersymmetric large $N_c$ QCD
in the strong 't Hooft coupling limit. 
We calculate the Euler-Heisenberg effective Lagrangian ${\cal L}(E)$,
which encodes the nonlinear response and the quantum decay rate
of the vacuum in a background electric field $E$, 
from the complex D-brane action in AdS/CFT.
We find that the decay rate given by Im$\,{\cal L}(E)$
becomes nonzero above a critical electric field
set by the confining force between quarks.
A large $E$ expansion of Im$\,{\cal L}(E)$ is found to coincide with that of
the Schwinger effects in QED, replacing its electron mass by the confining force.
Then, the time-dependent response of the system in a 
strong electric field is solved
non-perturbatively, and we observe a universal thermalization at a shortest 
timescale ``Planckian thermalization time" 
$\tau_{\rm th}\sim \frac{\hbar}{k_BT_{\rm eff}^\infty}
\sim
\frac{\hbar}{ k_B }E^{-1/2}$. 
Here, $T_{\rm eff}^\infty$ is 
an effective temperature which quarks feel in the nonequilibrium
state with nonzero electric current,
calculated in AdS/CFT as a Hawking temperature.
Stronger electric fields accelerate the thermalization,
and for a
realistic  value of the electric field  in RHIC experiment, 
we obtain $\tau_{\rm th}\sim 1~\mbox{[fm/c]}$, 
which is consistent with the believed value.
}

\preprint{
{\normalsize OU-HET-790} \\
{\normalsize RIKEN-MP-75}
}


\begin{document}
\setcounter{page}{1}

\section{Introduction}
\label{sec1}

Extreme environments, such as a strong electric field, is one of the
frontiers to test physical systems and to reveal new physical phenomena.
Particle physics is not an exception. 
The physics of quantum fields in strong 
external electric fields, {\it i.e.},
``strong-field quantum field theory"
has a very long history which even dates back to the 
development era of QED. 
Nevertheless, the dynamics of quantum fields and their vacuum 
in strong electromagnetic fields
has not been understood well yet, both theoretically and experimentally.

One of the present frontiers of strong field QFT
is to understand the instability of 
strongly interacting systems such as the confining vacuum in QCD. 
These theoretical researches are motivated by many experimental and 
natural phenomena. 
Famous examples are heavy ion collisions at RHIC
and LHC experiments where strong magnetic and electric fields are expected
to be instantly produced \cite{Kharzeev:2007jp}\footnote{ 
 See Ref.~\cite{YagiHatsudaMiake2005}
for a review on the quark gluon plasma. In this paper
we concentrate on purely Maxwell electric effect on QCD-like gauge theories, and will not consider
color-electromagnetic forces related to color glass condensate
\cite{Gelis:2010nm} 
and glasma \cite{Lappi:2006fp}.
}
Other examples include magnetors, neutron stars
whose surface magnetic field is extremely strong and would be related
to the dense hadron/quark phase inside the stars.

A particular interest is a relation between the confinement in QCD and
the strong electric field. Because quarks have electric charges, a 
strong electric field can induce a vacuum decay at which pairs of
a quark and an antiquark are produced from the vacuum to cancel
the background electric field. However to estimate the threshold critical electric field,
as well as to describe the physical decay process, is a difficult problem,
because of several reasons; first, QCD is strongly coupled so the standard 
perturbative
calculation does not work at low energy, and second, strong electromagnetic fields induces effective multi-photon vertices resulting in a complicated 
nonlinear electromagnetic effective action. 

\FIGURE[ht]{ 
\includegraphics[width=5cm]{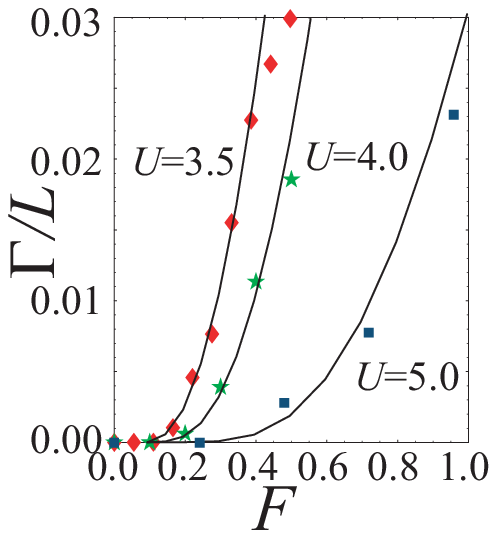}
\caption{
The ground state decay rate of the 1-dimensional Mott insulator
obtained numerically (dots) and analytically (solid lines) in the 
Hubbard model for several values of its interaction strength $U$ \cite{Oka2005b}. 
}
\label{fig:Mott}
}
Another source of our motivation comes from nonequilibrium phenomena in
strongly correlated electrons in condensed matter physics \cite{Oka_LZreview,Oka2003,Oka2005b,Eckstein2010c,MitraTakeiKimMillis06,Mitra2008,Dalidovich04,Karch2011,GreenSondhi05}. 
The physics of strong electric field breaking the vacuum is nothing but a famous notion
known as dielectric breakdown in material science. 
The Mott insulator, which hosts high Tc 
superconductivity when doped (=carriers are added) in cuprates, is a state in which the motion of electrons are frozen due to strong Coulomb repulsion. 
In order to realize ultrafast control of the Mott metal-insulator transition,
as well as superconductivity, its nonequilibrium response in
strong electric field is being intensively studied. 
An example of the decay rate of the Mott insulating 
ground state is shown in Fig.~\ref{fig:Mott}, 
and in fact one can observe the threshold behavior
with a critical electric field.

The renowned method for analyzing strongly coupled system, such as QCD,
is the 
AdS/CFT correspondence 
(also called holography or gauge/gravity duality) 
\cite{Maldacena:1997re,Gubser:1998bc,Witten:1998qj}. This
is a well-developed tool in 
string theory which enables us to analyze strongly coupled QCD analytically.
In this paper, we apply the gauge/gravity duality to a certain strongly coupled 
QCD-like gauge theory, and analyze the instability caused by a strong electric field.
The theory is a toy model of QCD --- 
${\cal N}=2$ supersymmetric large $N_c$ 1-flavor QCD.
Using the gauge/gravity duality, we present our analytic calculations on 
how the instability occurs under the strong electric field, for both
gapped and gapless cases (massive and massless quarks respectively), 
and for both time-independent 
and time-dependent nonequilibrium processes.

The main result of the present paper is an analytic computation of the full electromagnetic
effective Lagrangian for a strongly coupled QCD-like gauge theory. In particular,
the imaginary part of the effective Lagrangian shows the decay rate of the vacuum
of the gauge theory. 
To explain the notion in more detail, 
here let us briefly review the case of QED.

Shortly after Dirac's theory of positron,
Heisenberg and Euler \cite{Heisenberg1936}
as well as Weisskopf \cite{Weisskipf1936}
showed that the Dirac vacuum behaves as a dielectric, 
which can be polarized by electric fields. 
This effect can be elegantly captured by 
the effective Lagrangian first studied by 
Heisenberg and Euler \cite{Heisenberg1936}
and later by Schwinger \cite{Schwinger1951}
(see Ref.~\cite{Dunne2004,Dittrich}
for a review with recent developments
and Ref.~\cite{Oka_LZreview} for application in condensed matter physics.).
As for QED, the one-loop effective Lagrangian is given by 
\begin{eqnarray}
&&\mathcal{L}_{\rm QED}^{\rm 1-loop}=-i\ln\int\mathcal{D}[\psi,\bar{\psi}]
\exp\left[i\int d^4x \bar{\psi}(i\Slash{D}-m)\psi\right]/V
\label{eq:LdefQED}
\\
&&
=-i\ln\det(i\Slash{D}-m)/V\, ,
\end{eqnarray}
where the Dirac operator is $\Slash{D}=\gamma^\mu(\pa_\mu+ieA_\mu)$,
$A_\mu$ is a fixed external gauge potential with 
its field strength tensor $F_{\mu\nu}=\pa_\mu A_\nu-\pa_\nu A_\mu$,
$m$ the electron mass and $V$ is the spacetime volume.
The effective Lagrangian serves as a generating function
for non-linear responses of the vacuum, 
{\it e.g.}, vacuum polarization, light-light scattering.
The polarization induced by the static electric field $E$ is 
related to the real part of the effective Lagrangian
\begin{eqnarray}
P(E)=\frac{\pa \mbox{Re}\mathcal{L}}{\pa E}\, .
\label{eq:PE}
\end{eqnarray}
An interesting observation is that the effective Lagrangian 
has an imaginary part
\begin{eqnarray}
\mathcal{L}=\mbox{Re}\mathcal{L}+i\Gamma/2 \, .
\end{eqnarray}
The effective Lagrangian is related to the 
vacuum-to-vacuum amplitude via \cite{Oka2005b}
\begin{eqnarray}
\bra 0|U(t)|0\ket=e^{i\mathcal{L}vt},
\label{eq:020}
\end{eqnarray}
where $v$ is the spatial volume, $U(t)$ is a 
time-evolution operator with external fields, $|0\ket$ is the 
original vacuum state  (with no external fields).
Thus, $\Gamma$ gives the vacuum decay rate. 
For QED, the imaginary part is given, up to 1-loop order, by
\begin{eqnarray}
&&
\mbox{Im}\; \mathcal{L}_{\rm spinor}^{\rm 1-loop}
=\frac{e^2E^2}{8\pi^3}\sum_{n=1}^\infty\frac{1}{n^2}
\exp\left(-\frac{\pi m^2}{eE}n\right),
\label{eq:sch1}
\\
&&
\mbox{Im}\; \mathcal{L}_{\rm scalar}^{\rm 1-loop}
=\frac{e^2E^2}{16\pi^3}\sum_{n=1}^\infty\frac{(-1)^{n-1}}{n^2}
\exp\left(-\frac{\pi m^2}{eE}n\right)
\label{eq:sch2}
\end{eqnarray}
for spinor and scalar particles. 
The term $\exp\left(-\frac{\pi m^2}{eE}\right)$
represents a single quantum tunneling process
where a pair of an electron and a hole (positron) is created from the vacuum. 
The threshold field, often denoted as the Schwinger limit,
is given by
\begin{eqnarray}
E_{\rm cr}\sim \frac{m^2c^3}{e\hbar}\sim 10^{16} \; {\rm V/cm}.
\end{eqnarray}
Experimentally, it is still not easy to produce an electric field as strong
as the Schwinger limit even with the strongest lasers available.

In QCD, we expect a similar threshold electric field, since to liberate the
quark and the antiquark from a meson bound state one needs an electric field
which overwhelms the confining force existing between the quark and the antiquark.
A similar phenomena was studied in a toy-model of confinement, namely the 
massive Schwinger model (1+1 dimensional QED), where
the electric flux line between the electron and the positron gives a linear potential due to dimensionality
providing a confining force. It was shown that when 
the external electric field equal the confining force, 
a confinement-to-deconfinement transition takes place, with 
massless excitation described by Coleman's half-asymptotic state
\cite{Coleman1976}.
Thus, it is an interesting question to study QCD near and above the 
critical field. 

In this paper, we confine ourselves to the case of 1-flavor and at large $N_c$ 
for the gauge group $SU(N_c)$, 
instead of the realistic QCD, to study 
this liberation issue. The reason is that in QCD we still have gauge-singlet
observable states which are charged --- charged mesons and baryons. If we consider the 1-flavor and the large $N_c$, there exists no charged meson, 
and any baryon is infinitely heavy so it is difficult to create a pair of a baryon 
and an anti-baryon. 

The simplest example that can accommodate our interest is 
the ${\cal N}=2$ supersymmetric QCD whose gravity dual description 
\cite{Karch:2002sh} has been
studied in detail in various literature. In this paper, we shall use this supersymmetric QCD as an example to investigate the instability
of the confining strongly-coupled system.

In string theory, it is predicted that pair creation
of open strings should take place in the presence of 
electric fields
\cite{Bachas1992a,Bachas:1995kx}.
In the gauge/gravity setup, this leads to the pair creation
of quark antiquarks via the {\it holographic Schwinger effect}
studied by Semenoff and Zarembo
\cite{Semenoff2011}.
However this Schwinger effect treats an instanton process 
which is valid at small electric field, while we are interested in 
strong electric fields.

The Euler-Heisenberg effective Lagrangian is defined
from a partition function 
of a QFT in background electric fields $E$, {\it c.f.}, (\ref{eq:LdefQED}),
\begin{eqnarray}
{\cal L} =-i\ln Z_{\rm QFT}[E]/V\, .
\end{eqnarray}
We notice that the D-brane action, the low energy effective Lagrangian
of open strings on the D-brane, is already nonlinear Dirac-Born-Infeld
type \cite{Fradkin1985} and it serves as an Euler-Heisenberg Lagrangian
of our supersymmetric QCD via the gauge/gravity duality. 
The D-brane actions are valid also for
strong electric fields, so it suits our aim. More precisely, 
according to the gauge/gravity duality, {\it the Euler-Heisenberg 
effective Lagrangian of the ${\cal N}=2$ supersymmetric SU($N_c$)
QCD at the strong coupling limit and the large $N_c$ limit, is simply given by
the flavor D-brane effective action in a higher-dimensional curved spacetime. }
When we turn on the electric field, the effective Lagrangian 
shows a nonlinear dependence on the field. 
Furthermore, we observe that when the electric field exceeds
a critical value, it automatically induces an imaginary part in the D-brane 
action.\footnote{
For an early study of the imaginary part for a non-supersymmetric
confining gauge theory in AdS/CFT, see \cite{Kim:2008zn}.
} 
This gives a  signal of the instability of the strongly coupled QCD. 

\FIGURE[t]{ 
\includegraphics[width=15cm]{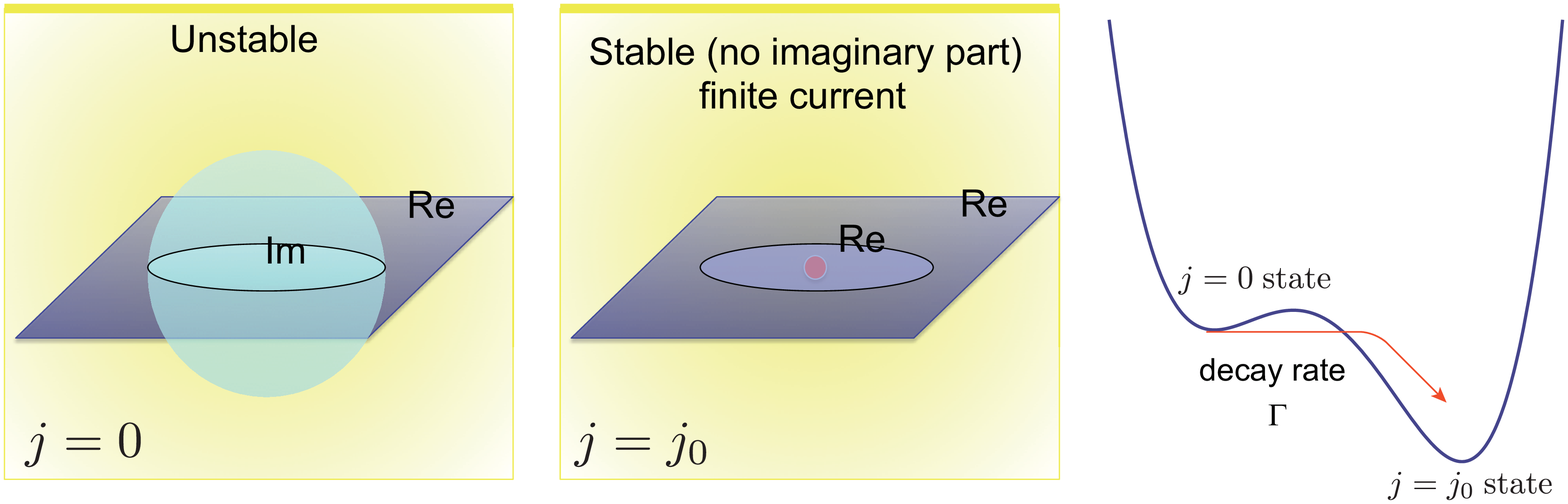}
\vspace{-10mm}
\caption{A schematic picture of the D7-brane in the $AdS_5\times S^5$ geometry.
The flat surface in the figure is the D7-brane on which we have a constant electric field.
Left: When the field exceeds a critical value, 
the initial zero current solution has an effective Lagrangian
with an imaginary part. This comes from a region in the D7 brane 
worldvolume near the D3 brane  (denoted by the blue sphere). 
Middle: When the stable solution with non-zero current $j=j_0$
is realized, the D7-brane effective action becomes real.
At the same time, an apparent horizon is formed. 
Left: Schematic stability diagram, where the electric field
makes the initial $j=0$ state unstable due to quantum tunneling.  
The system relaxes to the steady state with non-zero current. 
}
\label{fig:brane}
}
A schematic view of the D-brane configuration is given in Fig.~\ref{fig:brane}.
The electromagnetic effect takes place through the charged matter field (quarks)
whose dynamics is dictated by the flavor probe D-brane in the $AdS_5\times S^5$ 
geometry. 
Once we apply the constant electric field, and when the field 
exceeds a critical value, a region in the D-brane
worldvolume appears which 
gives an imaginary contribution to its effective action (the left panel of Fig.~\ref{fig:brane}). After the vacuum breakdown, the system should flow to
an nonequilibrium steady state with a constant electric current 
 (the central panel of
Fig.~\ref{fig:brane}). 
The nonequilibrium steady state has been studied in previous 
literatures  using the gauge/gravity duality. 
Our standpoint in this paper lies in the dynamical process 
starting from the unstable state in the 
left panel of Fig.~\ref{fig:brane}.

In this paper, using the gauge/gravity duality, we find the following; 
For the static situation, 
\begin{itemize}
\item We obtain the Euler-Heisenberg Lagrangian.
\item We find a critical electric field to have the imaginary part, 
which coincides with the confining force between quarks.
\item Surprisingly, the imaginary Lagrangian agrees with that of 
the Schwinger's calculation for QED in the strong-field expansion, 
via a simple re-interpretation of the 
electron mass by the confining potential.
\end{itemize}
In addition, since the gauge/gravity duality works also at
time-dependent situations, we can study the real time dynamics when 
a time-dependent electric field is applied. For this dynamical case,
\begin{itemize}
\item We find induced time-dependent electric current and
also its relaxation to a stationary current.
\item The system thermalizes with a universal Planckian time
$\tau_{\rm th}\sim \frac{\hbar}{k_BT_{\rm eff}^\infty}
\sim
\frac{\hbar}{ k_B }E^{-1/2}$ 
after an  abrupt introduction of the electric field. 
\end{itemize}
The quantum quench on the flavor (quark)  degrees of freedom has been studied in
the context of AdS/CFT \cite{Das:2010yw,Hashimoto:2010wv}, 
in addition to the thermalization due to
the quantum quench on the bulk gluons (and the spatial volume) in
AdS/CFT (see for example Refs.~\cite{Danielsson:1999fa,Bhattacharyya:2009uu,Janik:2006gp,Ebrahim:2010ra,AbajoArrastia:2010yt,Balasubramanian:2010ce,Buchel:2012gw,Heller:2012km,Balasubramanian:2013rva,Buchel:2013lla}). 
Recently, a universal time-scale of abrupt change in parameters in
AdS/CFT was reported \cite{Buchel:2013gba}.
Our universal Planckian 
thermalization is for a flavor quench with an abrupt changes in the electric fields.

The organization of this paper is as follows. In section 2, we briefly review
the D3-D7 brane configuration in the gauge/gravity duality
of the ${\cal N}=2$ supersymmetric large $N_c$ QCD,
and review the steady state solution with non-zero electric current in electric fields.
In section 3, we treat the gapless (massless) QCD. We 
evaluate the imaginary part of the Euler-Heisenberg action, and 
see the agreement with the standard result of the QED Schwinger effect.
In section 4, we study the system with a mass gap: the massive QCD.
Again we see agreement with the Schwinger effect, for large electric field
expansion. Section 5 is for the time-dependent analysis, with dynamical
electric current and the thermalization of the system seen in
the apparent horizon in the gravity dual.
The final section is devoted for discussions. In particular, a relation to the
previously-known holographic generalization of the Schwinger effect \cite{Semenoff2011}\footnote{
Recent analyses of the holographic Schwinger effect
include Refs.~\cite{Ambjorn:2011wz,Bolognesi:2012gr,Sato:2013pxa,Sato:2013iua,Sato:2013dwa}.
}
is discussed.


\section{Review: $I$-$V$ curve in holography}

\FIGURE[r]{ 
\hspace{2mm}
\includegraphics[width=6.5cm]{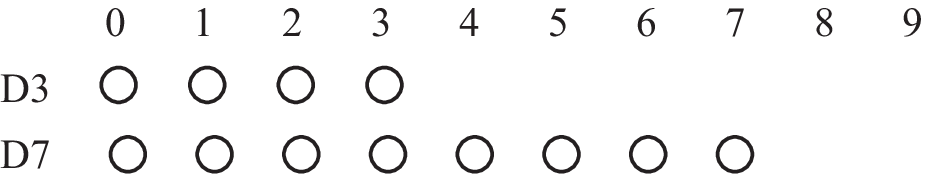}
\caption{Configuration of the D3-D7 system in 10 spacetime dimensions. 
}
\label{config}
}

Using gauge/gravity duality, one can analyze the full $E$-dependence 
of the current. Interestingly, in the
steady state, the relation between the current $j$ and the applied 
electric field $E$ is determined by requiring that the D-brane action is real
\cite{Karch:2007pd}.\footnote{
See also Ref.~\cite{Albash:2007bq} for a phase diagram 
and the metal-insulator transition. See Ref.~\cite{Erdmenger:2007cm} for a review
of the meson sector of the D3-D7 model. For confining gauge theories, 
analogous analyses have been made in 
Refs.~\cite{Bergman:2008sg,Johnson:2008vna,Kim:2008zn}. 
}
Our paper concerns the imaginary part of the D-brane action (which is related to the instability) on the other hand.
In this section, after presenting some basics of the gravity setup, 
we explain the reality condition which is used to realize the stable state.

In string theory, the simplest and best-understood system which has a charged fermion in the AdS/CFT set-ups is the
renowned D3D7 system \cite{Karch:2002sh}. The gauge theory realized at low energy is a supersymmetric QCD-like theory. 
Precisely, the content of the fields consists of ${\cal N}=4$ supersymmetric $SU(N_c)$
Yang-Mills theory in accompany with a ${\cal N}=2$ hypermultiplet in the fundamental representation of the 
color $SU(N_c)$ gauge group, in 1+3 dimensional spacetime. 
We may consider a finite temperature system of the gauge theory. 
The gravity dual of the gauge theory is a D7-brane put in the AdS black hole background.

The background metric is produced by a near horizon geometry of $N_c$ D3-branes put at a finite temperature,
\begin{eqnarray}
ds^2 = \frac{R^2}{z^2}
\left[
-\left(1-\frac{z^4}{z_{\rm H}^4}\right)dt^2 + \left(1-\frac{z^4}{z_{\rm H}^4}\right)^{-1} dz^2
+ d\vec{x}^2
\right] + R^2 d\Omega_5^2 \, .
\end{eqnarray}
The coordinate $z$ measures the holographic direction, and $z=0$ corresponds to the AdS boundary.
$z=z_{\rm H}$ is the horizon of the black hole, so the bulk spacetime is bounded as $0<z<z_{\rm H}$.
$R$ is the AdS radius, and 
$d\vec{x}^2 = dx_1^2 + dx_2^2+dx_3^2$.
The temperature $T$ and the gauge coupling constant $g_{\rm QCD}$ 
(or in other words the 't Hooft coupling constant $\lambda \equiv N_c g_{\rm QCD}^2$) are
 given in terms of the geometry as
\begin{eqnarray}
z_{\rm H} = \frac{1}{\pi T}\, , \quad R^4 = 4\pi g_s N_c \alpha'^2\, , \quad
2 \pi g_s = g_{\rm QCD}^2 \, .
\end{eqnarray}
Here $\alpha'$ is a string theory parameter given in such a way that $1/(2\pi\alpha')$ is the fundamental 
string tension, but 
will cancel out in any final results once we calculate gauge theory quantities. 

The D7-brane is placed to introduce the hypermultiplet (the charged fermions). For simplicity 
here we consider a massless
hypermultiplet (In the next section 
we study a massive system). The hypermultiplet comes from
oscillations of a string connecting the D3-brane and the D7-brane, so the massless-ness 
corresponds to the D7-brane configuration touching the stack of D3-branes.
The D7-brane action is
\begin{eqnarray}
S_{\rm D7} = - \mu_7 \int\! dt d^3\vec{x} dz d\Omega_3 \sqrt{-\det 
\left[
P[g]_{ab} + 2\pi \alpha' F_{ab}
\right]}\, .
\end{eqnarray}
Here $F_{\rm ab}$, the electromagnetic field on the probe D7-brane, 
plays the most important role in our paper. 
The external electric field determines the configuration of $F_{ab}$ 
and also the electric current in the gauge
theory. $\mu_7\equiv 1/((2\pi)^7 g_s \alpha'^4)$ is the D7-brane tension, and the integration measure $dzd\Omega_3$
is for the extra 4-dimensional space which differs from the D3-brane worldvolume (our 1+3 dimensional spacetime).  
$P[g]_{ab}$ is the induced metric on the D7-brane.

The local gauge field appearing here in the D7-brane action in the gravity side 
corresponds to the global symmetry of the fermion number in the gauge theory side. 
If we call the fermions "quarks",  this symmetry 
is related to the quark number (which is roughly a baryon number) 
in the supersymmetric QCD. Our aim is to introduce an 
``electric field" coupled to the charge of the quark number symmetry.\footnote{The fluctuation modes of the gauge fields on the D7-branes correspond to meson modes in the gauge theory side.}

For a flat D7-brane in the bulk geometry (which always solve the equations of motion for the scalar field on the D7-brane determining the shape of the D7-brane), the effective Lagrangian (which is the action divided by the 
spacetime volume) can be evaluated as
\begin{eqnarray}
{\cal L} &=& -2\pi^2 \mu_7
\int \! dz \;
\frac{R^8}{z^5}
\sqrt{\xi(F_{01},F_{0z},F_{1z})}\, , 
\label{eq:lagdef}
\\
\xi &\equiv& 
1-\frac{(2\pi\alpha')^2z^4}{R^4}
\left[
F_{01}^2 \left(1-\frac{z^4}{z_{\rm H}^4}\right)^{-1} 
+ F_{0z}^2 - F_{1z}^2 \left(1-\frac{z^4}{z_{\rm H}^4}\right) 
\right]\, .
\label{eq:xidef}
\end{eqnarray}
Note that here and in the following we write spacetime Lagrangian density rather than the action.
The factor $2\pi^2=$ Vol($S^3$) results from the $d\Omega_3$ integral.  

Let us see how the external electric field gives the electric current in this system. We are interested in a homogeneous configuration in the 
3-dimensional space in the gauge theory side, and thus, 
we simply put $\partial_x=0$. Then 
the equations of motion are
\begin{eqnarray}
&&
\pa_z\!\left[\frac{F_{0z}}{z\sqrt{\xi}}\right]=0 \, , \qquad
\pa_0\!\left[\frac{F_{0z}}{z\sqrt{\xi}}\right]=0\, , \; \nonumber \\
&&
\pa_z\!\left[\left(1\!-\!\frac{z^4}{z_{\rm H}^4}\right)\frac{F_{1z}}{z\sqrt{\xi}}\right]
+\pa_0\!\left[\left(1\!-\!\frac{z^4}{z_{\rm H}^4}\right)^{\!\!\!-1}
\!\!\!\frac{F_{01}}{z\sqrt{\xi}}\right]=0
\, .\quad
\label{eom3}
\end{eqnarray}
In particular when the field configuration is time-independent, we put $\partial_0=0$ and obtain 
\begin{eqnarray}
\partial_z \left[
\frac{F_{1z}}{z\sqrt{\xi}} \left(1-\frac{z^4}{z_{\rm H}^4}\right)  
\right] = 0 \, , \quad
\partial_z \left(
\frac{F_{0z}}{z\sqrt{\xi}} 
\right) = 0 \, .
\label{eq:der}
\end{eqnarray}
Note that a configuration of a constant field strength, $F_{01}=$ constant, with 
$F_{1z}=F_{0z}=0$, solves the equations of motion.

One can obtain a generic nontrivial solution, since from the equations 
(\ref{eq:der}), two integration constants $j$ and $d$ can be found,
\begin{eqnarray}
j = \frac{2\pi\alpha' F_{1z}}{z\sqrt{\xi}} \left(1-\frac{z^4}{z_{\rm H}^4}\right)  
 \, , \quad
d = \frac{2\pi \alpha' F_{0z}}{z\sqrt{\xi}} 
\, . 
\label{eq:integ}
\end{eqnarray}
Up to some normalization, $j$ is the electric current, and $d$ is the charge density. 
The solution can be explicitly written as
\begin{eqnarray}
&&
A_0
= \mu - \int^z_0  \!\!dz \;\; \frac{d}{2\pi\alpha'} z\sqrt{\xi} \, ,
\\
&&
A_1 = E t 
- \int^z_0  \!\!dz \;\; \frac{j}{2\pi\alpha'} z\sqrt{\xi}\left(1-\frac{z^4}{z_{\rm H}^4}\right)^{-1}  \, ,
\\
&&
A_z = 0 \, .
\end{eqnarray}
Here, $E=F_{01}$ is a constant electric field. In these expressions, $\xi$ is
given by 
\begin{eqnarray}
\xi = 
\frac{
1-\frac{(2\pi\alpha')^2z^4}{R^4}
E^2 \left(1-\frac{z^4}{z_{\rm H}^4}\right)^{-1} 
}
{1 + \frac{z^6}{R^4}
\left\{
d^2-j^2\left(1-\frac{z^4}{z_{\rm H}^4}\right)^{-1} 
\right\}
}
\, .
\label{eq:xiex}
\end{eqnarray}
This was obtained by substituting (\ref{eq:integ}) into the original definition of $\xi$ 
(\ref{eq:xidef}) and solve it for $\xi$.

\FIGURE[ht]{
\includegraphics[width=7cm]{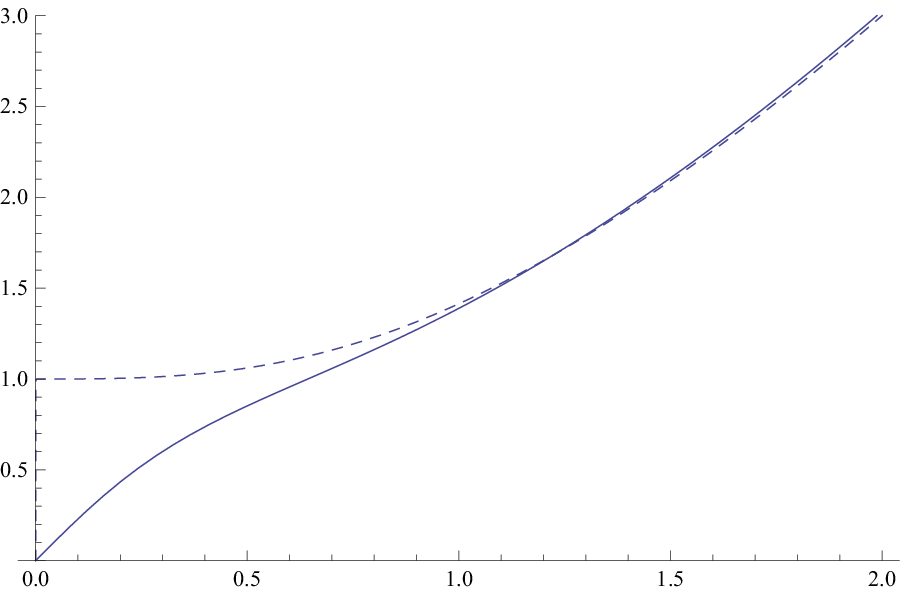}
\hspace{5mm}
\includegraphics[width=7cm]{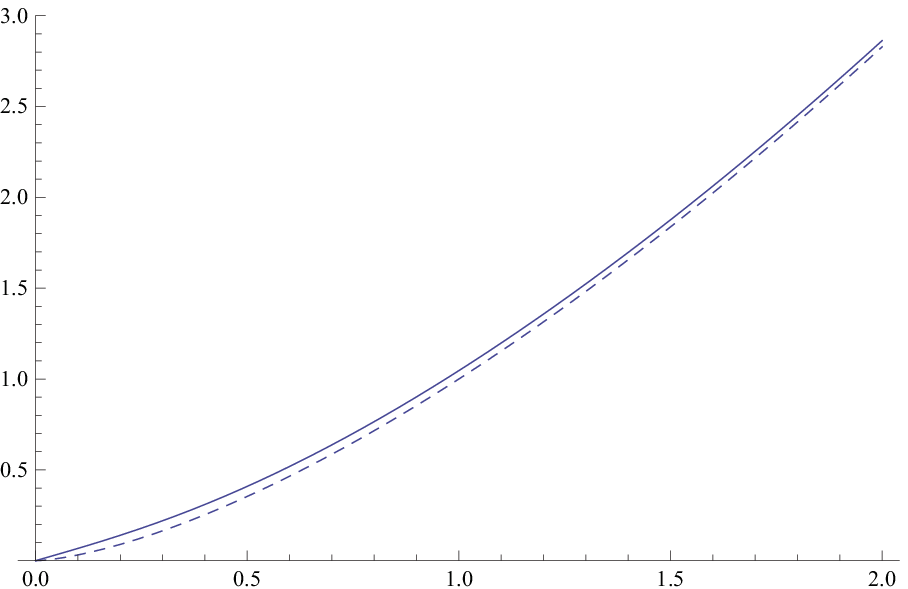}
\put(-140,-10){Electric field $E \; (d=0)$}
\put(-360,-10){Electric field $E \; (d\neq0)$}
\put(-180,100){$j_0$}
\put(-400,100){$j_0$}
\caption{
$IV$-characteristics of our massless supersymmetric QCD, given by eqn.(\ref{eq:iv-ch}).
The left (right) corresponds to a doped $d=1$ (non-doped $d=0$)
system. The dashed lines are for zero temperature $z_{\rm H}=\infty$, while the solid lines
are at a finite temperature $z_{\rm H}=1.5$.
We have set $R=1$ and $2\pi\alpha' =1$.}
\label{fig:iv}
}

It seems that $j$ can be determined irrespective of the electric field $E$, but it is not the case. Interestingly, the relation between $E$ and $j$ comes from the reality condition of the D-brane action \cite{Karch:2007pd,Karch:2007br,Erdmenger:2007bn}.\footnote{
The reality condition is reminiscent of that for a fundamental string in the $AdS_5$ 
black hole, for computing a quark drag force \cite{Herzog:2006gh,Gubser:2006bz}.
} 
Stable configuration of D-branes should not
admit an imaginary part of the action. However, in generic choice of $E$, $j$ and $d$, the quantity $\xi$ given by (\ref{eq:xiex}) can be negative, 
and since the action has a factor $\sqrt{\xi}$,
it leads to an imaginary part. So, any stable
configuration demands that $\xi$ should not be negative. 
This means that at a certain $z$ the denominator changes its sign 
and at the same $z$ the numerator should change its sign. 
Denoting this position $z$ as $z_p$, 
$z_p$ should solve two equations
\begin{eqnarray}
&&1 + \frac{z_p^6}{R^4}
\left\{
d^2-j^2\left(1-\frac{z_p^4}{z_{\rm H}^4}\right)^{-1} 
\right\} = 0 \, , \\
&&1-\frac{(2\pi\alpha')^2z_p^4}{R^4}
F_{01}^2 \left(1-\frac{z_p^4}{z_{\rm H}^4}\right)^{-1} = 0 \, .
\label{eq:solve2}
\end{eqnarray}
Eliminating $z_p$, we obtain the stable current $j=j_0$
written in terms of the electric field $E$,
the charge density $d$ and the temperature $T$ as
\begin{eqnarray}
j_0 = \sqrt{d^2 + \frac{\left((2\pi\alpha' E)^2 + R^4/z_{\rm H}^4\right)^{3/2}}{R^2}}
\frac{2\pi\alpha' E}{\sqrt{(2\pi\alpha' E)^2 + R^4/z_{\rm H}^4}}\, .
\label{eq:iv-ch}
\end{eqnarray}
This is the $I$-$V$ curve of our massless supersymmetric QCD
which is shown in Fig.~\ref{fig:iv}.

As a reference, we summarize two limits for this expression. 
When the density is zero ($d=0$), we have
\begin{eqnarray}
j_0 = \frac{2\pi\alpha' E}{R z_{\rm H}} 
\left(R^4 + (2\pi\alpha' E)^2 z_{\rm H}^4\right)^{1/4} \, .
\label{eq:d0iv}
\end{eqnarray}
When the temperature is zero ($z_{\rm H}=\infty$), we have
\begin{eqnarray}
j_0 = \sqrt{d^2 + \frac{(2\pi\alpha' E)^3}{R^2}} \, .
\label{eq:jT0}
\end{eqnarray}

We shall argue the meaning of the $I$-$V$ curve (\ref{eq:iv-ch}). 
In terms of the resistivity $\rho\equiv E/j_0$, our $I$-$V$ curve (\ref{eq:iv-ch}) 
gives the following relation, at the linear response $E\sim 0$, 
\begin{eqnarray}
\rho \sim \frac{T^2}{\sqrt{d^2 + T^6}}\, .
\label{eq:resi}
\end{eqnarray}
Here in (\ref{eq:resi}), to discuss a physical interpretation, 
we have ignored all numerical coefficients and the 
$\lambda$- and $N_c$- dependence. Let us see the temperature dependence for fixed $d\neq 0$ (doped). 
Our resistivity (\ref{eq:resi}) shows an
interesting behavior: 
At low temperature, it is a metal with a Fermi liquid
like behavior ($\rho \sim T^2$), while at high temperature it exhibits an insulator-type resistivity 
$d\rho/dT < 0$. This is interesting in two viewpoints. 
First, in holographic setups obtaining a Fermi-liquid 
is non-trivial since the system is in the strongly interacting limit
and the fermion propagators usually show abnormal behaviors.
Second, in most materials, the high temperature side of the phase diagram shows 
metallic physics while non-Fermi liquid is found in the low temperature side. 
This is opposite to what we observe. 
One possible reason for this comes from the fact 
that our system is not a pure Fermion 
system but also consists of strongly coupled bosons; 
In the high temperature regime, the bosons become highly excited which 
may disturb the transport of the fermions. 
The case with zero doping $d=0$ shows a very peculiar
strange metal behavior with $\rho\sim 1/T$ \cite{Karch:2007pd}.

\FIGURE[ht]{
\includegraphics[width=7cm]{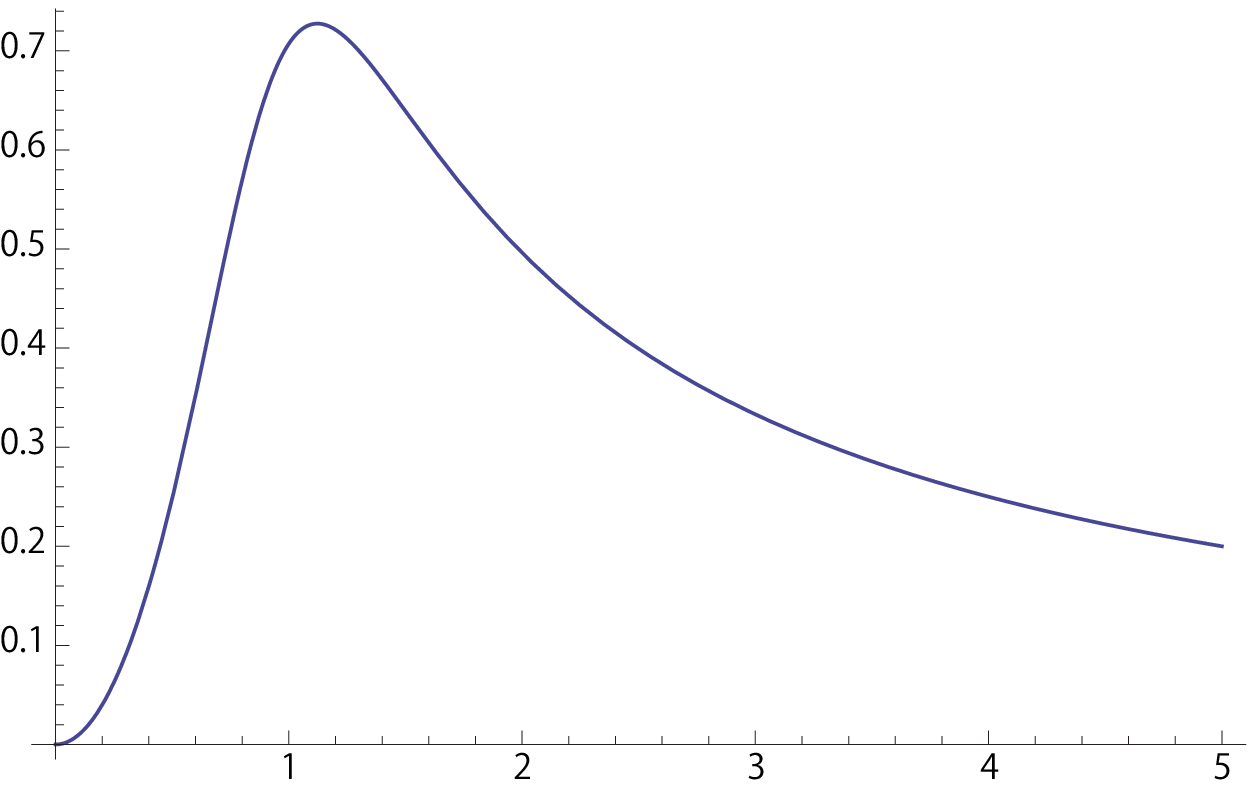}
\hspace{5mm}
\includegraphics[width=5cm]{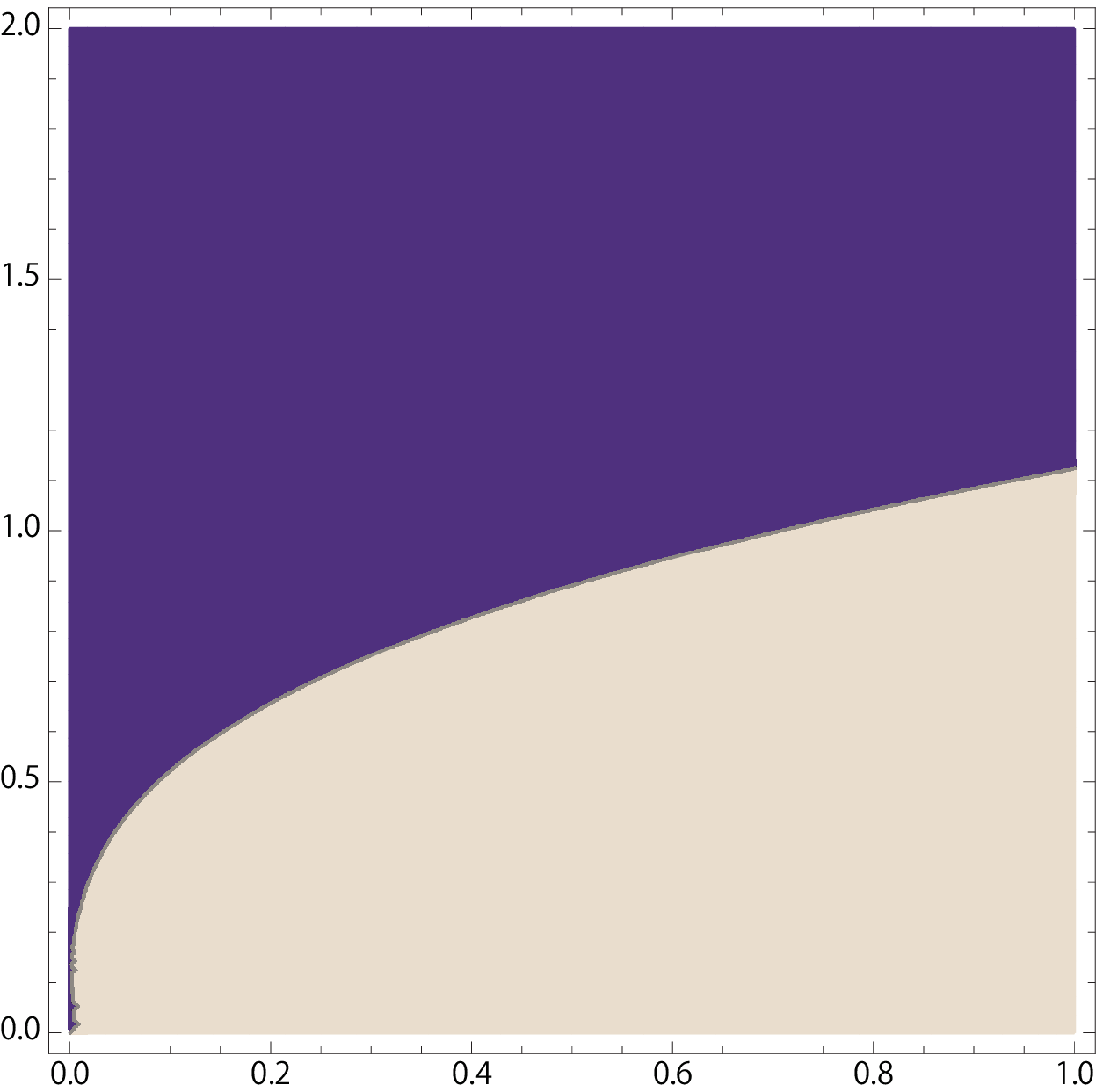}
\put(-40,-10){$T$}
\put(-260,-10){$T$}
\put(-160,110){$d$}
\put(-380,100){$\rho$}
\caption{Left: The resistivity (\ref{eq:resi}) at a finite charge density as a function of $T$. We have taken $d=1$. Right: The plot of $d\rho/dT$, the dark region gives 
an insulator-like phase $d\rho/dT<0$, while the light region is a metallic phase $d\rho/dT > 0$. At low temperature for fixed density $d$, we have Fermi-liquid like behavior $\rho \sim T^2$. }
\label{Fermi}
}

In the section, we have reviewed the $I$-$V$ curve of the ${\cal N}=2$ supersymmetric QCD, at finite quark density. The reality condition imposed on
the D-brane action determines the current $j$ as a function of the electric field $E$.


\section{Vacuum instability in a holographic gapless system}

Let us consider a sudden application of an electric field to the system.
Then it is expected that the system, which has been the vacuum, 
would become unstable. The instability 
is measured by the imaginary part of the effective action.

In the previous section, the $I$-$V$ curve of the supersymmetric QCD 
was derived by imposing the reality condition of the D-brane action. 
In this section, we relax this condition and study the 
physical meaning of the imaginary part. 
We treat
a gapless system (massless QCD) first, and in the next section we generalize
our result to a gapped system (massive QCD).

The imaginary part in the D-brane effective Lagrangian signals some instability.
In our case, it is natural to expect that the instability is related to the
existence of a more stable configuration; a phase with a constant electric current
at which the effective Lagrangian is no more imaginary.
In the past, the imaginary part of D-brane actions 
has not been
a subject of intensive study. However, D-brane actions are nothing but an open string
partition function (as seen from an off-shell formulation of string theory --- boundary string field theory), and it has been discussed that open strings in a background 
electric field develops instability associated with tachyonic modes
\cite{Burgess:1986dw,Nesterenko:1989pz,Bachas1992a,Bachas:1995kx}. 
Here our idea is to look into the imaginary 
 part of the flavor D-brane action in the gauge/gravity duality, seriously.


\subsection{The effective Lagrangian}
When the electric field is suddenly turned on at $t=0$,
the original vacuum solution given by $j=0$ becomes 
unstable in the presence of an electric field, 
and the reality condition which 
we considered in the previous section cannot be met. 
Let us investigate this simplest situation.
Let us put $j=0$ in the D7-brane effective Lagrangian \footnote{
The terminology ``Lagrangian" is used since 
this function describes the dynamics of the electromagnetic field 
which is coupled to the QCD degrees of freedom. 
The QCD degrees of freedom is integrated away and is 
effectively described by the (log of the) partition 
function calculated via holography, which is represented as
a $z$ integral.
}
\begin{eqnarray}
{\cal L} &=& -2\pi^2 \mu_7
\int_0^{z_{\rm H}} \!\!\!\! dz \;
\frac{R^8}{z^5}
\sqrt{
\left[1 + \frac{z^6d^2}{R^4}
\right]^{-1} 
\left[1-\frac{(2\pi\alpha')^2z^4}{R^4}
E^2 \left(1-\frac{z^4}{z_{\rm H}^4}\right)^{-1} 
\right]
} \, .
\label{eq:lagd7a}
\end{eqnarray}
From the square root in the expression, 
we notice that the integrand is real for small $z$, but at a certain $z_*$, it 
changes from a real function to an imaginary function:  
\begin{eqnarray}
z = z_* \equiv \alpha z_{\rm H}\, , \quad
\alpha \equiv  \left(1+\frac{(2\pi\alpha')^2 z_{\rm H}^4}{R^4}E^2 \right)^{-1/4}\, .
\end{eqnarray}
Since $z_* < z_{\rm H}$ for nonzero $E$ and nonzero $z_{\rm H}$ (the vanishing
$z_{\rm H}$ means infinite temperature $T=\infty$),
there always exists imaginary ${\cal L}$.

As briefly explained in the introduction, 
Fig.~\ref{fig:brane} illustrates the brane configuration and the regions of the brane
which give the real and the imaginary parts of the D-brane action. The flat sheet 
denotes the D7-brane in the $AdS_5\times S^5$ geometry.
When there is no current, the central region of the D7-brane (indicated by the inside of a sphere whose radius is $z_*$), 
given by $z>z_*$, contributes to the imaginary part. The outer region, closer to the
boundary of the $AdS_5$, contributes to the real part of the D7-brane effective Lagrangian through the integration.
On the other hand, when there exists the current, we can make all the
region to be giving a real number for the D7-brane action, as we have reviewed in
the previous section.

In terms of $z_*$, the effective Lagrangian (\ref{eq:lagd7a}) 
simplifies to the following expression,
\begin{eqnarray}
{\cal L} &=& -2\pi^2 \mu_7 R^6 \sqrt{R^4 + (2\pi\alpha')^2 E^2 z_{\rm H}^4}
\int_0^{z_{\rm H}} \!\!\!\! dz \;
\frac{1}{z^5}
\sqrt{
\frac{z_*^4 - z^4}{
(1+d^2z^6/R^6)(z_{\rm H}^4-z^4)}
} \, .
\end{eqnarray}
Defining $y \equiv z/z_{\rm H}$, we obtain explicit integral formulas
for the real and the imaginary parts of the effective Lagrangian,
\begin{eqnarray}
{\rm Re}{\cal L} &=& -2\pi^2 \mu_7 \frac{R^8}{z_{\rm H}^4 \alpha^2} 
\int_0^\alpha \!\!\!\! dy \;
\frac{1}{y^5}
\sqrt{
\frac{\alpha^4 - y^4}{
(1+d^2z_{\rm H}^6 y^6 /R^6)(1-y^4)}
\label{eq:ReRes}} \, , 
\\
{\rm Im}{\cal L} &=& 2\pi^2 \mu_7 \frac{R^8}{z_{\rm H}^4 \alpha^2} 
\int_\alpha^1 \!\!\!\! dy \;
\frac{1}{y^5}
\sqrt{
\frac{y^4-\alpha^4}{
(1+d^2z_{\rm H}^6 y^6 /R^6)(1-y^4)}
} \, .
\label{eq:ImRes}
\end{eqnarray}
Here, the electric field dependence is included in $\alpha$, 
and the temperature in 
$z_{\rm H}$. 
The imaginary part (\ref{eq:ImRes})\footnote{The sign of the imaginary part 
is chosen to be positive among $-1= (\pm i)^2$.} characterizes 
the vacuum instability of
the ${\cal N}=2$ supersymmetric QCD, at the large $N_c$ and at 
the strong coupling limit.

Fig.~\ref{fig:imL} is a plot of the imaginary part of the effective Lagrangian (\ref{eq:ImRes}) as a function of $E$. Obviously, it is a monotonic function of
$E$, and it agrees with the intuition that larger values of $E$ would induce
stronger instability. The plot also shows that in doped ($d\neq 0$) systems the
imaginary part increases as the temperature increases. This again is 
consistent with the intuition that there are more decay channels in 
systems with higher 
temperature to the current flowing state. The zero-temperature case is analyzed in App.\ref{app:T0}.

\FIGURE[ht]{
\includegraphics[width=7cm]{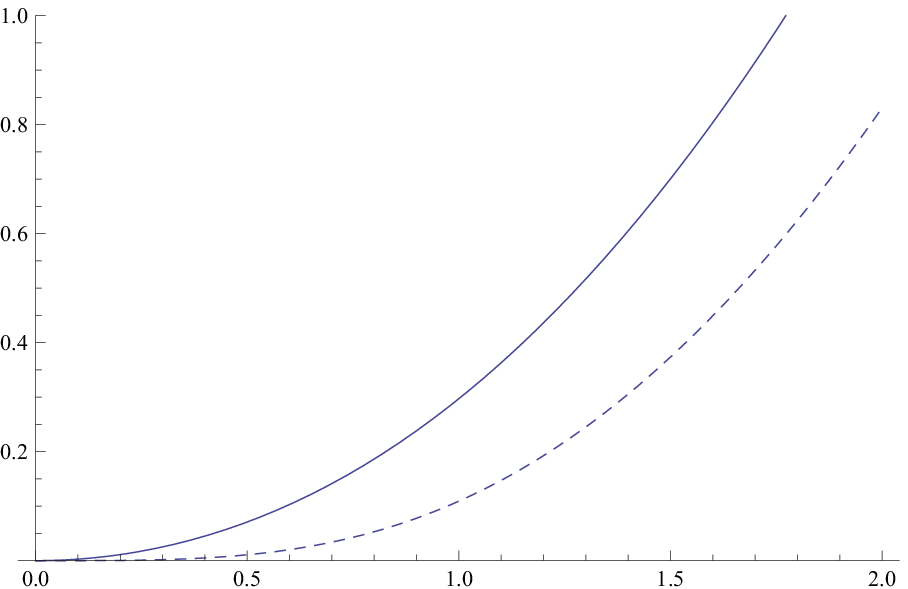}
\hspace{5mm}
\includegraphics[width=7cm]{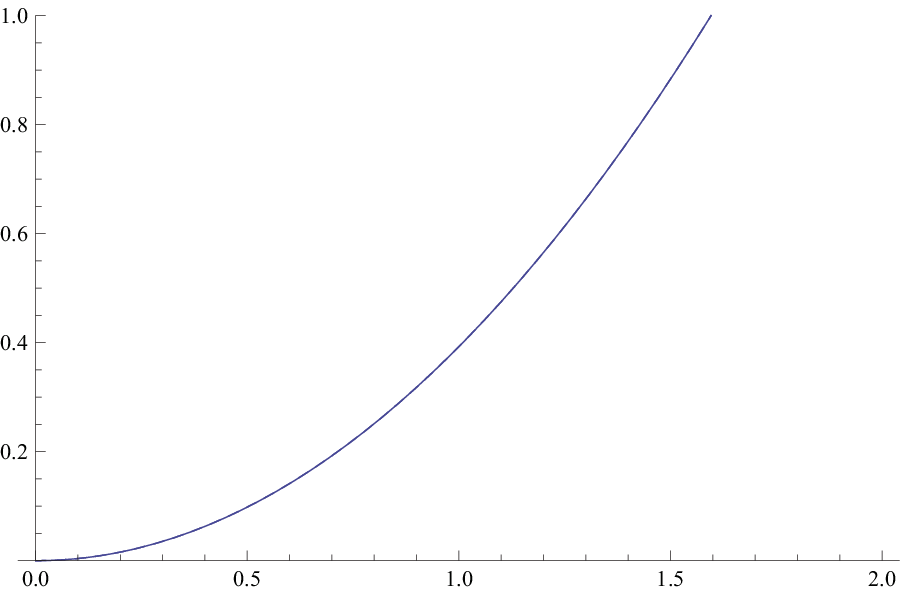}
\put(-140,-10){Electric field $E \; (d=0)$}
\put(-360,-10){Electric field $E \; (d\neq0)$}
\put(-180,100){Im ${\cal L}$}
\put(-400,100){Im ${\cal L}$}
\caption{
The imaginary part of the lagrangian of our massless supersymmetric QCD, given by 
(\ref{eq:ImRes}).
The left (right) corresponds to a doped $d=1$ (non-doped $d=0$)
system. The dashed lines are for zero temperature $z_{\rm H}=\infty$, while the solid lines
are at a finite temperature $z_{\rm H}=1$.
We have set $R=1$ and $2\pi\alpha' =1$. For no doping ($d=0$), there is no temperature dependence.}
\label{fig:imL}
}


\subsection{Coincidence with Schwinger at vanishing density}

The imaginary part of the effective Lagrangian (\ref{eq:ImRes}) is a function of 
the electric field $E$, the temperature $T$ and the charge density $d$. 
Let us consider the case of no doping, $d=0$. Using the formula
\begin{eqnarray}
\int_\alpha^1 \!\!\!\! dy \;
\frac{1}{y^5}
\sqrt{
\frac{y^4-\alpha^4}{
1-y^4}
} = \frac{\pi}{8}\frac{1-\alpha^4}{\alpha^2}\, ,
\end{eqnarray}
the imaginary part (\ref{eq:ImRes}) is rewritten as
\begin{eqnarray}
{\rm Im}\, {\cal L}\Bigm|_{d=0}  &=&  \frac{\pi^3}{4} \mu_7  R^4 (2\pi\alpha')^2E^2 \, .
\end{eqnarray}
We use the string theory expressions for the AdS radius $R$ and the D7-brane tension $\mu_7$ as
\begin{eqnarray}
R^4 = (2\pi\alpha')^2\frac{\lambda}{2\pi^2}\, , \quad
\mu_7 = \frac{1}{(2\pi)^7 g_s \alpha'^4} = \frac{1}{4\pi^2 g_{\rm SQCD}^2 (2\pi\alpha')^4}\, , 
\end{eqnarray}
then the imaginary part (\ref{eq:ImRes}) is written solely by $N_c$ and the electric field as
\begin{eqnarray}
{\rm Im}\,{\cal L}\Bigm|_{d=0} &=&  \frac{N_c}{32\pi} E^2 \, .
\label{eq:imld0}
\end{eqnarray}

Quite surprisingly, our result (\ref{eq:imld0}) coincides with 
the standard result for the QED vacuum instability by Schwinger \cite{Schwinger1951} and Weisskopf \cite{Weisskipf1936}
discussed in the introduction, {\it i.e.}, 
eqns.~(\ref{eq:sch1}) and (\ref{eq:sch2})\footnote{
Note that in the standard normalization of gauge fields
our $E$ is written as $eE$ where $e$ is the gauge coupling constant.}.
To compare them with our result, we 
take a massless limit $m=0$, and use formulas
 $\sum_{n=1}^{\infty} n^{-2} = \pi^2/6$,  $\sum_{n=1}^{\infty} (-1)^{n-1}n^{-2} = \pi^2/12$. Furthermore, since we are working in the ${\cal N}=2$ supersymmetric QCD, the charged matter fields consist of an ${\cal N}=2$ hypermultiplet 
 (a single Dirac fermion and two complex scalars). 
 It turns out that the total Schwinger effect is
\begin{eqnarray}
N_c \left({\rm Im}\;  {\cal L}_{\rm spinor}^{\rm 1-loop} \biggm|_{m_e=0} + 
2 \; {\rm Im}\;  {\cal L}_{\rm scalar}^{\rm 1-loop} \biggm|_{m_e=0}\right) = \frac{N_c}{32\pi} E^2 \, .
\label{eq:sch}
\end{eqnarray}
The factor $N_c$ is from the number of hypermultiplets 
(as they are in the fundamental representation of $SU(N_c)$ color symmetry). 
This one loop QED result of the Schwinger effect (\ref{eq:sch}) agrees with our imaginary part (\ref{eq:imld0}).

This agreement 
is somewhat unexpected: In our ${\cal N}=2$ supersymmetric QCD
we are working in the strong coupling limit
where the effective action should include information of all the summation
of the gluon-exchange diagrams, while the calculation of the one loop 
Euler-Heisenberg effective Lagrangian in QED (and the scalar QED) 
is done in the free limit with no photon nor gluon mediated interactions.  
 In addition, we have no 
supersymmetry once we turned on the electric field, so, BPS properties
which may often protect observables from perturbative gluonic 
corrections are not expected.
In the next section, we generalize our calculation to the massive case, and 
will see further agreement.

We have one more comment on our result (\ref{eq:imld0}), which is 
the independence from
the temperature. Although we are working with arbitrary temperature,
the result (\ref{eq:imld0}) does not depend on the temperature. 
This is natural from the Schwinger mechanism viewpoint, as follows. 
First, the instanton action of the electron in the background electric field
is made by a circular
worldline of the electron in Euclideanized spacetime whose radius is $m/E$, and
the effect of the temperature appears in the periodicity in the Euclideanized time direction. 
So if $2\pi/T > m/E$, the temperature should give 
no effect on the instanton action. Because we took the gapless limit $m=0$,
this inequality is always satisfied, so there should be no temperature-dependence in the Schwinger's result. 
Our temperature independence in (\ref{eq:imld0}) indicates that
this intuitive picture via instantons
may hold even at our strong coupling limit for gluons.

\subsection{Energy difference between the $j=0$ and $j=j_0$ states}

Let us look more closely at the real part of the Euler-Heisenberg Lagrangian 
(\ref{eq:ReRes}). When the background temperature is zero ($z_{\rm H}=\infty$),
the expression simplifies to
\begin{eqnarray}
{\rm Re}\, {\cal L} _{\rm unstable}= -2\pi^2\mu_7\frac{(2\pi\alpha' E)^2 R^4}{2}
\int_0^1 \! \frac{d\tilde{y}}{\tilde{y}^3} \sqrt{1-\tilde{y}^2} \, ,
\label{eq:enetot}
\end{eqnarray}
where we have shifted the integration variable as $\tilde{y}\equiv \alpha y$.
This corresponds to the energy of the total unstable system.

Now, suppose the system relaxes to another phase where we have a constant
electric current. This state is reviewed in section 2, and we expect that our 
instability shown by the imaginary part signals the decay 
to this state with the current. 
So, let us look at the energy difference between our unstable system
given by (\ref{eq:enetot}) and the energy of the stable state with the current.

When we have the electric current, the D-brane effective action is
given by $\sim \sqrt{\xi}$ where $\xi$ is given by (\ref{eq:xiex}). The effective
Lagrangian is easily evaluated, with $d=0$ and $j_0\neq 0$, with the integration
variable and boundary chosen in such a way that the Lagrangian is real,
\begin{eqnarray}
{\rm Re}\, {\cal L} _{\rm stable}= -2\pi^2\mu_7\frac{(2\pi\alpha' E)^2 R^4}{2}
\int_0^1 \! \frac{d\tilde{y}}{\tilde{y}^3} \sqrt{\frac{1-\tilde{y}^2}{1-\tilde{y}^3}} \, .
\label{eq:enetot2}
\end{eqnarray}

We are interested in the energy difference, and it can be calculated as
\begin{eqnarray}
{\rm Re}\, {\cal L} _{\rm unstable}-{\rm Re}\, {\cal L} _{\rm stable}= 
\frac{c}{8\pi^2} N_c E^2 
 \, ,
\label{eq:enetot3}
\end{eqnarray}
with a positive 
numerical constant $c \sim 0.855$. It is quite intriguing that the energy difference is actually finite, and the real part of our complex effective Lagrangian
certainly makes sense.\footnote{Readers may have suspected that there would
be a divergence in energy formulas. In fact there appears a divergence
but it can be removed by a holographic renormalization \cite{Bianchi:2001kw,Karch:2005ms}. Here we are evaluating
the difference, so there is no need to perform explicitly the holographic renormalization. }


\section{Vacuum instability in holographic system with mass gap}

A more interesting setup is a strongly correlated system with a mass gap.
In this section, we consider the case with massive quarks. In the ${\cal N}=2$ supersymmetric QCD, we can turn on the mass for the hypermultiplet. The quark and the squark have the same mass $m_q$. In the holographic description,
this quark mass is translated to the distance between the D7-brane and the $N_c$ D3-branes. 

We work out the real part and the imaginary part of the D7-brane effective action in the $AdS_5 \times S^5$ spacetime, to obtain the full Euler-Heisenberg Lagrangian.
We find that the imaginary part develops only for $E$ larger than the critical
electric field $E_{\rm cr}$ set by the confining potential. This is what we expect from the example of $1+1$ dimensional QED, the massive Schwinger model.
We shall also see that the first two terms in the large $E$ expansion of the imaginary part of the Lagrangian coincides with the weak coupling Schwinger calculation, somehow unexpectedly.

\subsection{Critical electric field and the confining force}
Our goal here is to write down the D7-brane action with finite external electric fields for a massive system. 
The massive version of our D3/D7 system is described by a geometry where 
the D7-brane is separated from the center of the $AdS_5$ geometry.
In the absence of the electric field, the D7-brane is just a flat hyperplane \cite{Karch:2002sh}. So, we consider a flat D7-brane in the $AdS_5$, and
put a constant electric field $E$ on the D7-brane.\footnote{ 
Note that when the electric field is turned on 
 in the massive case, in order to analyze the D-brane time evolution,
the bending of the D7 brane described by the scalar field 
must be taken into account.  
This is in contrast to the massless case where the flat D7-brane solves the
equations of motion.
However, this difficulty does not appear in the calculation of the 
vacuum instability described by the effective Lagrangian: 
We only need the zero field solution, and not the full 
time dependent solution. 
This is because we we are not interested in the final stable configuration but only in the instability of the initial vacuum state. 
The Euler-Heisenberg Lagrangian captures the instability 
as well as the nonlinear properties of the vacuum. It is related to the 
vacuum persistence amplitude by $\langle 0 | U(t) | 0 \rangle 
=e^{i\mathcal{L}vt}$ ($v$: spatial volume),
where $|0\rangle$ is the vacuum without the electric field while $U(t)$ is the 
time-evolution matrix with the electric field.
The on-shell time evolution will be studied in section \ref{sec:td}.

}

Let us consider the quark mass $m_q$. The distance between the D7-brane
and the center of $AdS_5$ is given by $\eta = 2\pi\alpha' m_q$, as the fundamental string tension is given by ${\cal T}_{\rm F1} = 1/(2\pi\alpha')$. 
At  zero temperature, the background is just the $AdS_5\times S^5$ geometry,
and the D7-brane action with the constant electric field
$F_{0x}=E$ is given by
\begin{eqnarray}
{\cal L} = -2\pi^2 \mu_7
\int_0^\infty \! dz \;
\frac{R^8}{z^5}
\sqrt{
1-\frac{(2\pi\alpha')^2 R^4 }{(\eta^2 + R^2/z^2)} E^2} \, .
\label{eq:fullD7}
\end{eqnarray}
Obviously, putting $\eta=0$ brings it back to the previous case (\ref{eq:lagdef}).
$z$ is the radial coordinate, and $z=\infty$ is the boundary.
\begin{figure}[ht]
\centering 
\includegraphics[width=8.5cm]{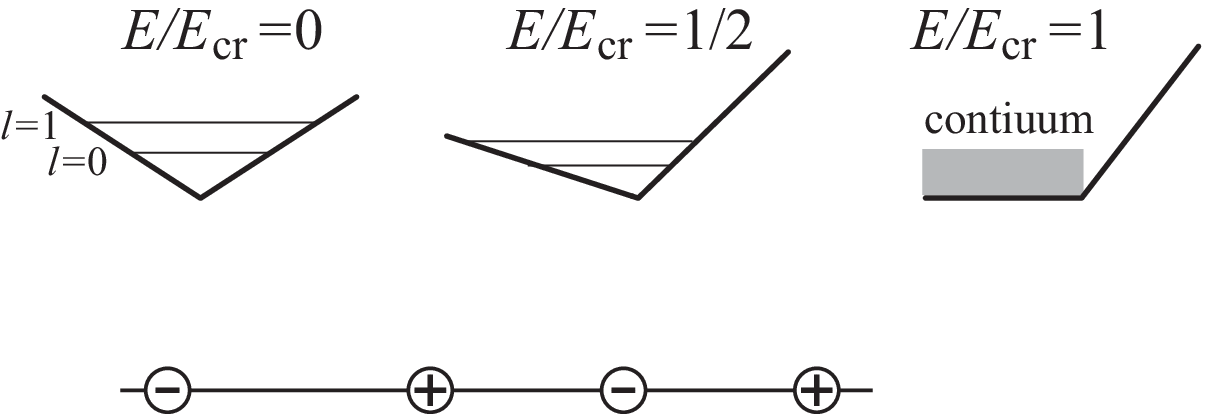}
\caption{
Above: Schematic confining potential of the quark and antiquark in the presence of an electric field.
At $E=E_{\rm cr}$, the potential becomes flat in one direction
inducing the liberation of the quark-antiquark pair.
This leads an emergent massless continuum states.  
Below: Coleman's half-asymptotic state realized at $E=E_{\rm cr}$
in the massive Schwinger model.
Quarks and antiquarks can freely move in the direction of the 
electric field as long as the charges are in a 
plus-minus-plus- $\cdots$  order.
}
\label{fig:Coleman}
\end{figure}

Now, as before, the factor inside the square root can be negative, 
which gives us
the imaginary part of the action. The square root becomes zero at
\begin{eqnarray}
z=z_0 \equiv \frac{R^2}{\sqrt{2\pi\alpha' R^2 E - \eta^2}} \, .
\end{eqnarray}
So, if this $z_0$ exists, we have the imaginary action. 
The condition for existence of the imaginary part is 
\begin{eqnarray}
E > E_{\rm cr} \equiv \frac{\eta^2}{2\pi\alpha' R^2} = \frac{\sqrt{2}\pi}{\sqrt{\lambda}} m_q^2\, .
\label{eq:Ecr}
\end{eqnarray}
We find that there is a critical electric field beyond which the action develops an imaginary value.

Now, the important lesson from the massive Schwinger model ($1+1$-dimensional QED with massive electrons) \cite{Coleman1976}
is that the originally confined charges become 
liberated when the external electric field acting on the 
electron positron pair balances the confining force 
between them (see Fig.~\ref{fig:Coleman}). 
So let us compare the critical electric field (\ref{eq:Ecr}) obtained  
from the D7-brane action and see whether it coincide with the confining force.
In Ref.~\cite{Kruczenski:2003be}, 
from the computation of small Wilson loop on the D7-brane, the QCD string tension at the meson sector of the D3D7 model was calculated. 
The quark antiquark potential for distance $L$ is given by
\begin{eqnarray}
V = \frac{\sqrt{2}\pi}{\sqrt{\lambda}} m^2 \; L \, .
\end{eqnarray}
Comparing it with (\ref{eq:Ecr}), the confining force exactly agrees with our critical electric field.

We confirmed the physical intuition that the instability of the system 
becomes present only when we have an electric field large enough to cancel the confining force.
This indicate that generically systems with confining force may have a similar 
behavior.


\subsection{Imaginary part: coincidence with Schwinger}

Let us examine the imaginary part of the D7-brane action
(\ref{eq:fullD7}). The imaginary part comes from the integral over the region 
$z_0 < z < \infty$, 
\begin{eqnarray}
{\rm Im}\, {\cal L} = 2\pi^2 \mu_7 (2\pi\alpha') R^{10} \sqrt{E^2-E_{\rm cr}^2} 
\int_{z_0}^\infty \! dz \;
\frac{\sqrt{(z^2-z_0^2)(z^2 + \tilde{z}_0^2)}}{z^5(z^2 \eta^2 + R^4)} \, ,
\label{eq:EH}
\end{eqnarray}
where we have defined 
\begin{eqnarray}
\tilde{z}_0\equiv \frac{R^2}{\sqrt{2\pi\alpha' R^2 E + \eta^2}} \, .
\end{eqnarray}

\FIGURE[tb]{
\includegraphics[width=7cm]{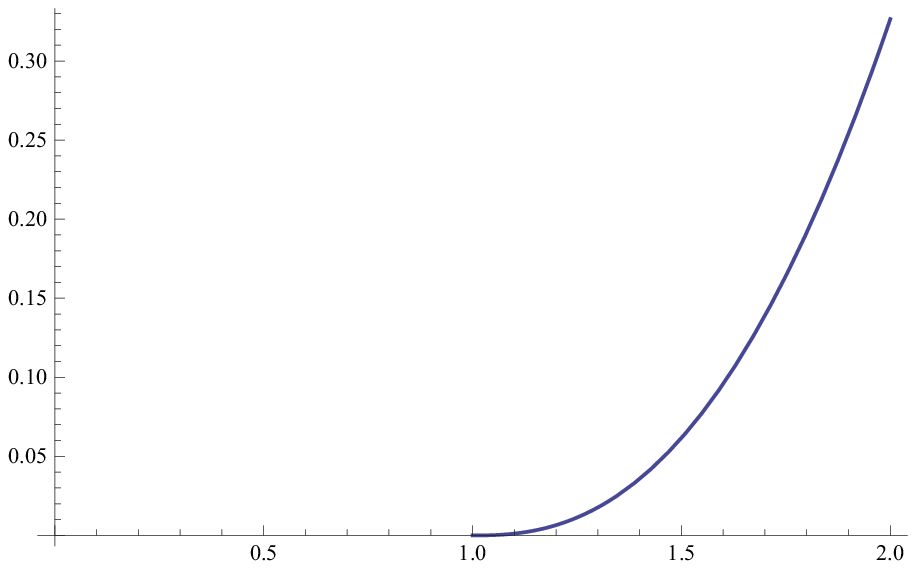}
\put(-40,-10){$E/E_{\rm cr}$}
\put(-180,100){Im ${\cal L}$}
\caption{
The imaginary part of the lagrangian of our massive supersymmetric QCD, 
given by (\ref{eq:imlfig}).
}
\label{iv}
}

In Fig.~\ref{iv}, this imaginary part (\ref{eq:EH}) is shown. A smooth increase of the
imaginary part, starting at $E=E_{\rm cr}$, is observed.

We want to compare our result with the QED results, {\it i.e.}, (\ref{eq:sch1}) and 
(\ref{eq:sch2}). 
In the massless QCD case studied in the previous section, 
we saw complete agreement of the imaginary part of the effective Lagrangian.
In the present massive case, since we have another mass scale, 
which is the quark mass, 
it is appropriate to consider the large $E$ expansion. 
The reason is that at small $E$, because of the confinement 
taking place below the energy scale $m^2/\sqrt{\lambda}$ in QCD, 
there should be no agreement between QED. 

To investigate the large $E$ behavior, let us define $\epsilon \equiv E_{\rm cr}/E$ 
and use a new radial coordinate 
\begin{eqnarray}
v \equiv \frac{z^2}{z_0^2}-1 \, .
\end{eqnarray}
Then the imaginary part (\ref{eq:EH}) is rewritten as
\begin{eqnarray}
{\rm Im}{\cal L} &=& \pi^2 \mu_7 (2\pi\alpha')^2 R^4 E_{\rm cr}^2 
\frac{(1-\epsilon)^{5/2}}{\epsilon^2}
\int_{0}^\infty \! dv \;
\frac{\sqrt{v(2+v+\epsilon v)}}{(v+1)^3 (1+\epsilon v)}
\nonumber 
\\ 
&=& 
\frac{N_c}{8\pi^2}
E_{\rm cr}^2 
\frac{(1-\epsilon)^{5/2}}{\epsilon^2}
\int_{0}^\infty \! dv \;
\frac{\sqrt{v(2+v+\epsilon v)}}{(v+1)^3 (1+\epsilon v)} \, .
\label{eq:imlfig}
\end{eqnarray}
Let us expand this D-brane result for a large $E$, {\it i.e.} the 
small $\epsilon$
expansion,
\begin{eqnarray}
\frac{(1-\epsilon)^{5/2}}{\epsilon^2}
\int_{0}^\infty \! dv \;
\frac{\sqrt{v(2+v+\epsilon v)}}{(v+1)^3 (1+\epsilon v)}
= \frac{\pi}{4} \frac{1}{\epsilon^2}
\left(
1 + \frac{4}{\pi} \epsilon \log \frac{\epsilon}{2} - \frac{1}{3\pi}\epsilon^3 + {\cal O}(\epsilon^4)
\right)
 \, .
\end{eqnarray}
In terms of the physical quantities, this is equivalent to the following expression
of the imaginary part,
\begin{eqnarray}
{\rm Im}{\cal L} =\frac{N_c}{32\pi} E^2
\left(
1 + \frac{4}{\pi} \frac{E_{\rm cr}}{E} \log \frac{E_{\rm cr}}{2E} 
- \frac{1}{3\pi}\left(\frac{E_{\rm cr}}{E}\right)^3 + {\cal O}\left(\left(\frac{E_{\rm cr}}{E}\right)^4\right)
\right) \, .
\label{eq:imexpm}
\end{eqnarray}
Of course for a massless quark limit $m=0$, in other words $E_{\rm cr}=0$, this expression reduces to the massless QCD case (\ref{eq:imld0}). Because of the
nonzero mass $m$, we find a sub-leading correction of the form 
$(E_{\rm cr}/E) \log (E_{\rm cr}/E)$.

Although there is no strong reason for us to expect that 
this supersymmetric QCD result should agree with the 
Schwinger's QED result, let us anyway compare them.
The QCD result is (\ref{eq:sch1}) and (\ref{eq:sch2}), which is in fact a sum
of dilogarithmic functions,
\begin{eqnarray}
\lefteqn{
N_c \left({\rm Im}\;  {\cal L}_{\rm spinor}^{\rm 1-loop}  + 
2 \; {\rm Im}\;  {\cal L}_{\rm scalar}^{\rm 1-loop} \right) 
}
\nonumber \\
&=&
 \frac{N_c}{8\pi^3} E^2 
\biggl[
{\rm Li}_2(\exp (-m_e^2\pi/E))-{\rm Li}_2(-\exp (-m_e^2\pi/E))
\biggr]
\, .
\end{eqnarray}
where ${\rm Li}_2(z) \equiv \sum_{n=1}^\infty z^n/n^2$ is the dilogarithm function.
Using the known expansion of the dilogarithm functions, we obtain
\begin{eqnarray}
{\rm Im} \, {\cal L}^{\rm 1-loop}_{N_c {\rm spinor}+2N_c {\rm scalar}} =\frac{N_c}{32\pi} E^2
\left(
1 + \frac{4}{\pi} \frac{m_e^2}{E} \log \frac{m_e^2\pi}{2eE} 
- \frac{1}{9\pi^2}\left(\!\frac{m_e^2 \pi}{E}\!\right)^3 
\!\!+\! {\cal O}\left(\!\left(\!\frac{m_e^2 \pi}{E}\!\right)^4\right)
\!\right) \, .
\nonumber\\
\end{eqnarray}
Note here that $e$ in this expression is the base of natural logarithms, 
not the electric charge.
We notice that at the nontrivial sub-leading order, this expression coincides with
the supersymmetric QCD at strong coupling, (\ref{eq:imexpm}), once we identify
\begin{eqnarray}
E_{\rm cr} \leftrightarrow m_e^2 \, .
\end{eqnarray}
This coincidence is quite nontrivial.
First, in the holographic QCD calculation, we have infinite gluon exchanges
while in QED of course there is no gluon. Second, although the QCD side has 
supersymmetries, the electric field breaks the supersymmetries completely, 
so we cannot generically expect any cancellation of the higher diagrams
via the supersymmetries.  Nevertheless we obtained interesting coincidence 
between the famous 1-loop QED results with our holographic 
calculation of the imaginary part of the effective action. 
One possible reason for this is due to 
some conformal symmetry in the gluon sector, but we leave the detailed research for a future work.

\subsection{Euler-Heisenberg Lagrangian}

So far, we focused on the imaginary part of the effective action. 
In the present 
subsection we will examine the real part. 

\subsubsection{Effective action below the critical electric field}

When $E\leq E_{\rm cr}$, the D7-brane action, {\it i.e.}, the Euler-Heisenberg action, remains real.
It is important to obtain the series expansion in $E$
since the coefficients become the 
``non-linear optical response function".
Using the expansion formula
\begin{eqnarray}
\sqrt{1-x} = 1 + \sum_{n=1}^\infty \frac{(2n-3)!!}{n!} \left(\frac{-x}{2}\right)^n \, ,
\end{eqnarray}
we can evaluate the integral in (\ref{eq:EH}) term by term after the expansion,
to obtain
\begin{eqnarray}
{\cal L} =
-2\pi^2 \mu_7 R^8
\left[
\int_{\delta z}^\infty \frac{dz}{z^5}
+\frac{(2\pi\alpha')^2}{4R^4}\left(1 + \log\frac{\eta^2 (\delta z)^2}{R^4}\right)E^2
\right.\qquad\qquad
\nonumber \\
\left.
\qquad+ 
\frac{\eta^4}{R^8}\sum_{n=2}^\infty \frac{(2n-3)!!}{4 (2n-1)(n-1)n!} 
\left(
\frac{-(2\pi\alpha')^2 R^4}{2 \eta^4}
\right)^n E^{2n}
\right] \, .
\end{eqnarray}
In this expression, The term $E^{2n}$ with $n=0$ and $n=1$ has a divergent
coefficient, so we included a UV cutoff $z=\delta z$, so that the integral is
for the region $\delta z < z < \infty$.  
Following the standard holographic renormalization technique \cite{Bianchi:2001kw,Karch:2005ms}, we
add a local counter term,
\begin{eqnarray}
\delta {\cal L} =
2\pi^2 \mu_7 R^8
\left[
\int_{\delta z}^\infty \frac{dz}{z^5}
+\frac{(2\pi\alpha')^2 E^2}{4R^4}\log\frac{\eta^2 (\delta z)^2}{R^4}
\right]
\end{eqnarray}
The coefficients (the finite term) are determined in such a way that the renormalized action does not have any $\alpha'$ dependence if written in terms of 
the gauge theory variables. After the subtraction, we finally arrive at the
expression for the Euler-Heisenberg Lagrangian,
\begin{eqnarray}
{\cal L} =
-\frac{N_c}{16\pi^2} E^2
-\frac{N_c \lambda}{8 \pi^4} E_{\rm cr}^2
\sum_{n=2}^\infty \frac{(2n-3)!!}{4 (2n-1)(n-1)n!} 
\left(
\frac{-E^2}{2 E_{\rm cr}^2}
\right)^n \, .
\end{eqnarray}
We have found an infinite tower of nonlinear responses in the electric field.
Note that the coefficient of the first term, which is a charge renormalization, has ambiguity as a renormalization scheme dependence.

\subsubsection{Effective action above the critical electric field}
When the electric field exceeds the critical field
$E>E_{\rm cr}$, the integral region that contributes to the 
real part of the effective action
is restricted to $0 < z < z_0$. The integral becomes
\begin{eqnarray}
{\rm Re}\, {\cal L} = -2\pi^2 \mu_7 (2\pi\alpha') R^{10} \sqrt{E^2-E_{\rm cr}^2} 
\int_0^{z_0} \! dz \;
\frac{\sqrt{(z_0^2-z^2)(z^2 + \tilde{z}_0^2)}}{z^5(z^2 \eta^2 + R^4)} \, .
\end{eqnarray}
Using a new coordinate $u\equiv z^2/z_0^2$,
we obtain
\begin{eqnarray}
{\rm Re}\, {\cal L} = -\pi^2 \mu_7 (2\pi\alpha')^2 R^4 E_{\rm cr}^2 
\frac{(1-\epsilon^2)^{5/2}}{\epsilon^2}
\int_0^1 \frac{du}{u^3} \;
\frac{\sqrt{(1-u)(u+1 + \epsilon(u-1))}}{1+\epsilon(u-1)}
 \, .
\end{eqnarray}
The integral is divergent, but with a UV cutoff $u=\delta$
it can be exactly calculated as
\begin{eqnarray}
\lefteqn{
\frac{(1-\epsilon^2)^{5/2}}{\epsilon^2}
\int_\delta^1 \frac{du}{u^3} \;
\frac{\sqrt{(1-u)(u+1 + \epsilon(u-1))}}{1+\epsilon(u-1)}
}
\nonumber \\
&&
=\frac{-\left(3 + \log 2\right)}{2} \epsilon^{-2}
+\frac{-\pi}{2} \epsilon^{-1}
-\frac12 - \frac12 \epsilon^{-2} \log \epsilon + I_{\rm div}
\end{eqnarray}
where
\begin{eqnarray}
I_{\rm div} \equiv 
\frac12 \left((\delta z)^2 \frac{\eta^2}{R^4}\right)^{-2}
+ \frac{1}{2\epsilon^2} \log \left[(\delta z)^2 \frac{\eta^2}{R^4}\right]
\, .
\end{eqnarray}
Here we have used the relation between the cutoff in $u$ and the cutoff in $z$,
\begin{eqnarray}
\delta = \frac{(\delta z)^2}{z_0^2} = (\delta z)^2 \frac{\eta^2}{R^4} \frac{1-\epsilon}{\epsilon}.
\end{eqnarray}
This $I_{\rm div}$ is subtracted from the full effective action, by following the
holographic renormalization (it is the same counter term as before.)

Finally, we obtain a finite real part of the Euler-Heisenberg Lagrangian, 
\begin{eqnarray}
{\rm Re}\, {\cal L} = \frac{N_c}{8\pi^2}
\left[
\frac{3 + \log 2}{2} E^2
+\frac{\pi}{2} E_{\rm cr} E
+\frac12 E_{\rm cr}^2 + \frac12 E^2 \log \frac{E_{\rm cr}}{E}
\right]
 \, .
\end{eqnarray}
Note that again the coefficient in front of the $E^2$ term
has an ambiguity coming from the renormalization scheme dependence.
The obtained Lagrangian has an interesting terms: the linear term $E$, and
the log term $E^2 \log E$. 
We do not understand the reason that the effective Lagrangian suddenly 
acquires a term that is of odd power in the electric fields. 
It is expected to be related to the nature of the quantum state that
appears at the critical field strength, with possible common aspects as  
Coleman's half asymptotic state \cite{Coleman1976}.

\section{Time-dependent process: relaxation and thermalization}
\label{sec:td}
Up to now, we have concentrated on the steady (un)stable state. 
One virtue of the gauge/gravity duality is that we can 
investigate  time-dependent nonlinear responses of the system. 
In this section, we perform a time-dependent 
calculation of the system in which the electric field is switched on from zero to nonzero.

We expect two phenomena accordingly. One is the excitation 
and relaxation of the system, and the other is thermalization. 
Both are quite interesting and nontrivial in strongly correlated systems
and are related to the physics studied experimentally at RHIC and LHC. 
Indeed we shall see in the following that
the system relaxes to a nonequilibrium steady state 
with a constant current flow which is ``thermalized".
The nonequilibrium thermal state is characterized by 
the {\it effective Hawking temperature $T_{\rm eff}$}
that we define below. 
This is calculated from the 
induced metric which depends on the D7-brane dynamics
and thus on the external electric field. 
The effective Hawking temperature relaxes to a 
steady state value, 
which is nothing but the standard Hawking temperature for a given
metric with a horizon, and through the gauge/gravity duality it is interpreted
as a Matsubara temperature of the gauge theory.\footnote{
Note that our effective temperature is the temperature felt by
fluctuation modes (mesons) on the D7-brane, and is different 
from the temperature of the background bulk geometry 
(the heat bath of gluons). In this section we set 
the bulk gluon temperature $T=0$ 
(which corresponds to $z_{\rm H}=\infty$ and the 
geometry is just the $AdS_5\times S^5$).}

We find a universal behavior in this relaxation process.
In fact, the time scale for the relaxation becomes
universal, {\it i.e.}, no matter how quickly we apply the constant electric field, the thermalization time takes a value given by the 
Planckian time scale $1/T_{\rm eff}$ with a parameter 
independent proportionality factor.

In this section, we first
explain the calculation scheme in holography and then 
present results of the thermalization.
We consider the massless case $\eta=0$ throughout this section. 


\subsection{Equations of motion and effective Hawking temperature}

The equations of motion for the massless case at zero temperature is
(\ref{eom3}) with $z_{\rm H}=\infty$. We choose a gauge $A_z=0$, then 
the equations of motion can be rewritten as follows:
\begin{eqnarray}
\pa_z\left(\frac{F_{0z}}{z\sqrt{\xi}}\right)=0
\, ,\quad
\pa_0\left(\frac{F_{0z}}{z\sqrt{\xi}}\right)=0
\, ,\quad 
-\pa_z\left(\frac{\pa_zA_1}{z\sqrt{\xi}}\right)
+\pa_0\left(\frac{\pa_0A_1}{z\sqrt{\xi}}\right)=0
\, .
\end{eqnarray}
>From the first two equations, 
we can set
\begin{eqnarray}
\frac{F_{0z}}{z\sqrt{\xi}}=d
\end{eqnarray}
using a constant $d$, which is nothing but the charge density, as we explained
in section 2. 
Since
$\xi=1-(z^4/R^4)\{
(\pa_z A_0)^2+(\pa_0 A_1)^2-
(\pa_z A_1)^2
\}$,  we have 
\begin{eqnarray}
\xi=\frac{
1-\frac{z^4}{R^4}\{
(\pa_0A_1)^2
-(\pa_zA_1)^2
\}
}{1+\frac{z^6}{R^4}d^2} \, .
\end{eqnarray}
Thus, the explicit equations of motion for the gauge field 
$A_1$ is given by
\begin{eqnarray}
-\pa_z\left(\frac{\sqrt{1+\frac{z^6}{R^4}d^2}\pa_zA_1}{z\sqrt{
1-\frac{z^4}{R^4}\{
(\pa_0A_1)^2
-(\pa_zA_1)^2\}}}\right)
+\pa_0\left(\frac{\sqrt{1+\frac{z^6}{R^4}d^2}\pa_0A_1}{z\sqrt{
1-\frac{z^4}{R^4}\{
(\pa_0A_1)^2
-(\pa_zA_1)^2\}}}\right)=0 \, .
\nonumber\\
\label{eq:eoma1}
\end{eqnarray}

We can devide $A_1$ into two parts, 
\begin{eqnarray}
A_x=-\int^tE(s)ds+h(t,z) \, ,
\end{eqnarray}
where $E(t)$ is the external electric field, while
$h(t,z)$ is a dynamical degree of 
freedom that satisfies
$h(t,z=0)=0,\;\pa_zh(t,z=0)=0$.
According to the standard AdS/CFT dictionary,
the higher derivative term corresponds to the current $j$.
To be precise,
$h$ is related to the current \cite{Karch:2007pd} by
\begin{eqnarray}
\pa_z^2h(t,z=0)=\bra J^x\ket/\mathcal{N} \, ,
\end{eqnarray}
where ${\cal N}$ is a normalization factor.

Let us define an effective temperature for mesons. 
In any nonequilibrium time-dependent process,
definition of the temperature is ambiguous. 
However, in the gravity dual part of
the gauge/gravity duality, it is natural to define the temperature 
as  an ``effective Hawking temperature" defined from the
inverse of the apparent horizon radius,
\begin{eqnarray}
T_{\rm eff}=\sqrt{\frac{3}{8}}\frac{1}{\pi z_{\rm AH}} \, .
\label{eq:Teff}
\end{eqnarray}
The apparent horizon converges to the event horizon once the system
converges to a steady state, and in that asymptotic 
limit the above definition coincides
with the standard Hawking temperature of a ``black hole" made by 
the induced metric on the probe D7-brane. 
For the derivation of (\ref{eq:Teff}), see Appendix B.

The location of the apparent horizon $z=z_{\rm AH}$ is defined as a time-dependent
quantity. We follow the standard definition of the apparent horizon; 
it is defined as a hyper surface whose volume does not change if shifted
along a null outward geodesics.
First, using the induced metric of the gauge fields on the 
D7-branes, the volume $V_{\rm AH}$ 
of a hypersphere with a fixed $t$ and a fixed radius $z$ becomes
\begin{eqnarray}
V_{\rm AH}
&=&\int d^3 d^3\theta^I
\sqrt{G_{11}G_{22}G_{33}
G_{I=1,I=1}G_{I=2,I=2}G_{I=3,I=3}}
\nonumber \\
&=&V_3\cdot\mbox{Vol}(S^3)\frac{R^6}{z^3}\mu_7^{-\frac{3}{2}}
\left\{1+\frac{z^4}{R^4}(2\pi\alpha')^2(-F_{01}^2-F_{0z}^2+F_{1z}^2)
\right\}^{3/4}
\nonumber \\
&&
\hspace{40mm}
\times
\left\{1+\frac{z^4}{R^4}(2\pi\alpha')^2(-F_{01}^2+F_{1z}^2)
\right\}^{1/2} \, .
\label{eq:volume}
\end{eqnarray}
Here $G_{ab}$ is the effective metric which the gauge fluctuation on the D7-brane
feels. The explicit expression is given in Appendix B.
The null vector $(v^t,v^z)$ in the $(t,z)$-slice satisfies
\begin{eqnarray}
G_{ab}v^av^b=0 \, .
\label{eq:null}
\end{eqnarray}
The ratio of the components is calculated as 
\begin{eqnarray}
v_t/
v_z=\frac{\frac{z^2}{R^2}2\pi\alpha'F_{1z}\frac{z^2}{R^2}2\pi\alpha'F_{01}+\sqrt{\xi}
}{-1+(\frac{z^2}{R^2}2\pi\alpha'F_{01})^2}.
\end{eqnarray}
Using this, the position of the apparent horizon $z=z_{\rm AH}(t)$ is defined by 
the solution of the equation
\begin{eqnarray}
\delta V_{\rm AH}(t,z_{\rm AH}(t))\equiv
(v_t\pa_t+v_z\pa_z)V_{\rm AH}(t,z)\biggm|_{z=z_{\rm AH}(t)}=0.
\label{eq:Vol}
\end{eqnarray}

The effective Hawking temperature $T_{\rm eff}(t)$ defined by eqn.~(\ref{eq:Teff})
is a function of the boundary time $t$. 
Suppose that the apparent horizon is at $(t_{\rm AH}, z_{\rm AH}(t_{\rm AH}))$, 
then the boundary time $t_{\rm b}$ 
for this point is given by 
\begin{eqnarray}
t_{\rm b}=t_{\rm AH}+z_{\rm AH}\, ,
\end{eqnarray}
because the information of the presence of the apparent horizon of the
brane needs to be propagated to the boundary via bulk geometry along the
bulk null geodesics $(1,-1)$.

In the steady state limit ($t\to \infty$), the 
effective temperature should converge to the asymptotic value
which is the Hawking temperature \cite{Kim:2011qh} given by 
\begin{eqnarray}
T_{\rm eff}^\infty=\sqrt{\frac{3}{8}}\frac{1}{\pi z_p} \,.
\label{eq:effHaw}
\end{eqnarray}
This is because the apparent horizon approaches the event horizon
given by $z=z_p$ on the D7-brane, c.f., (\ref{eq:solve2}), in the long time limit.

\subsection{Planckian thermalization accelerated by electric fields}

We are ready to analyze explicitly the time-dependence of the system 
as a response to the change of the background electric field. 
At time $t=0$, we smoothly switch on the electric field to its static value,
given by (Fig.~\ref{currentT1} (a))
\begin{eqnarray}
E(t)=\frac{E}{2}(1+\tanh(\omega t)) \, .
\label{eq:timeE}
\end{eqnarray}
The time length of the ramp is given by $1/\omega$. 
Here $E$ is the constant electric field,
so after a sufficient time, the system should relax to a nonequilibrium steady
state with a constant current flow, which was reviewed in section 2.
In numerical calculations, we use the normalization $2\pi\alpha'=1$.

\FIGURE[hbt]{
\includegraphics[width=15cm]{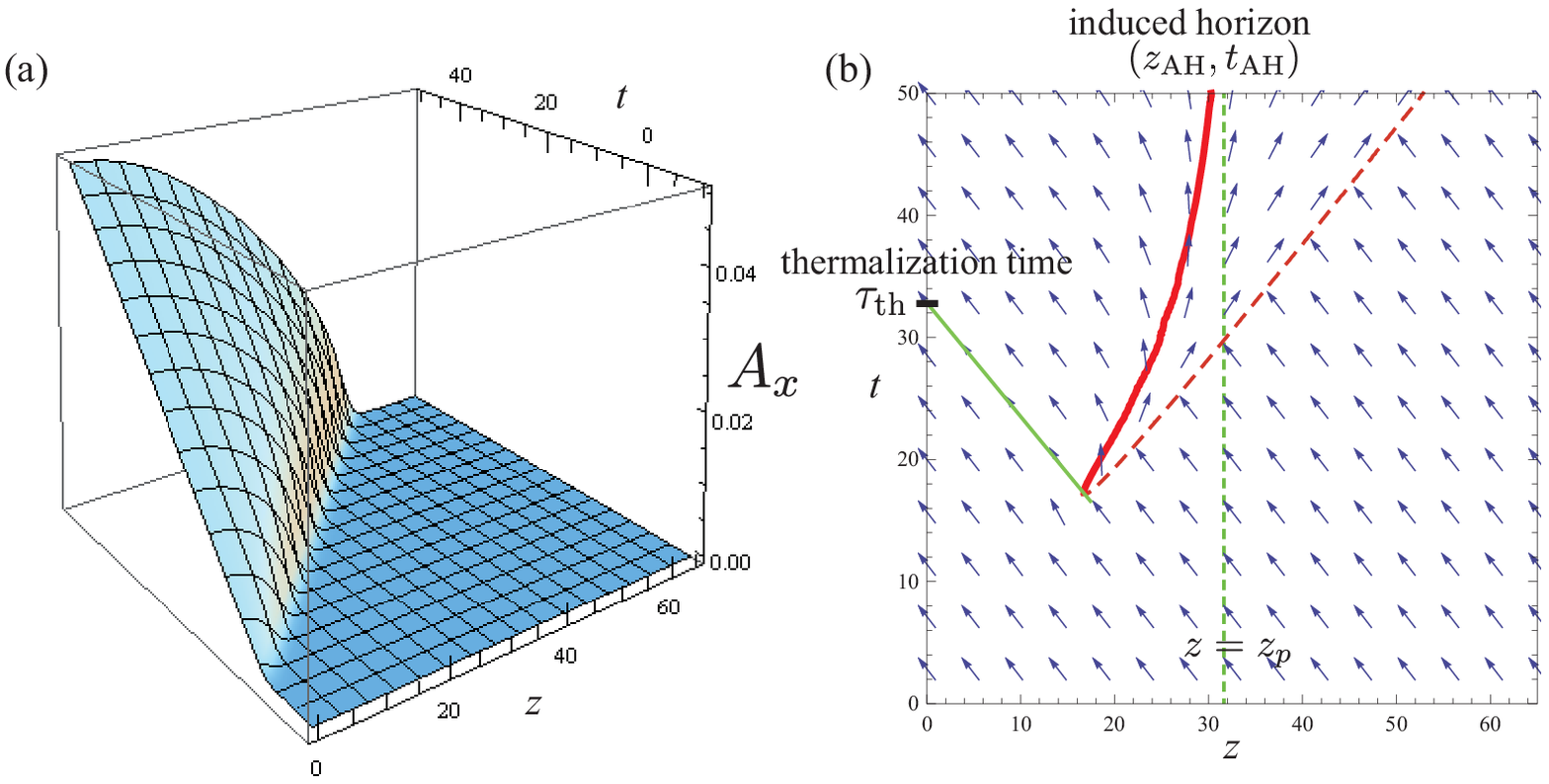}
\caption{
(a) The gauge field $A_x(z,t)$ for $E=0.001$
in the undoped case ($d=0$)
calculated with parameters $T=0,\;\omega=0.7,\;R=1$.
(b) The induced horizon (red solid line).
In the long time limit,
it converges to the steady state value $z=z_p$ denoted by the 
green dotted line. 
The red (solid and dashed) lines are determined 
from the condition $\delta V_{\rm AH}=0$ given in (\ref{eq:Vol}).
The vector field plots the direction of the null vector
$(v_z,v_t)/\sqrt{v_z^2+v_t^2}$. 
}
\label{horizon1}
}

Figure \ref{horizon1} (a) plots the gauge field $A_1 = A_x(z,t)$
obtained by solving the equations of motion (\ref{eq:eoma1}). 
After the electric field is switched on around $t=0$, 
$A_x$ begins to increase and this effect travels inward into the AdS direction. An apparent horizon is induced, which is determined by the
condition (\ref{eq:Vol}); Two branches of solutions exist
as plotted as the solid and dashed red lines in Fig.~\ref{horizon1} (b).
The apparent horizon corresponds to the branch
that is closer to the boundary (solid line), which we denote by $(z_{\rm AH},t_{\rm AH})$. 
This is confirmed by plotting the null vector direction 
(blue vector field) which points straight-upward (no $z$-component)
on the horizon. 
The formation of the horizon starts to affect the physics 
of the boundary only after its earliest boundary time 
(minimum of $t_{\rm b}=t_{\rm AH}+z_{\rm AH}$) is reached. 
We call this time  the ``thermalization time" $\tau_{\rm th}$.

\FIGURE[ht]{
\includegraphics[width=16cm]{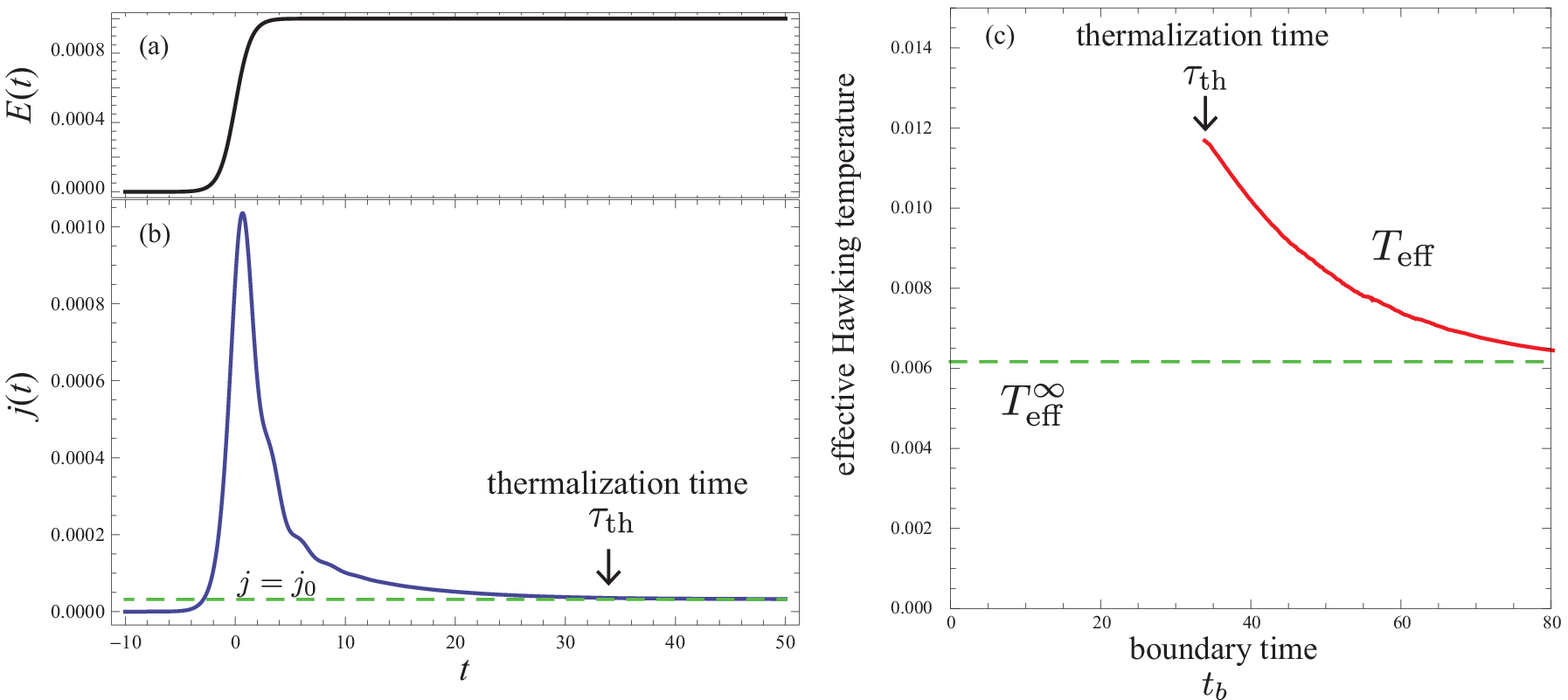}
\caption{(a) The electric field $E(t)$.
(b) Induced current $j(t)$. It converges to the 
steady state value $j=j_0$ (\ref{eq:iv-ch}).
(c) The effective Hawking temperature (\ref{eq:Teff})
plotted against the boundary time. 
The ``thermalization time" is defined as the boundary time
of the horizon formation. 
In the long time limit, $T_{\rm eff}$ converges
to the steady state value $T_{\rm eff}^\infty$.
Same parameters as Fig.~\ref{horizon1} is used.
}
\label{currentT1}
}
The time evolution of the current is shown in Fig.~\ref{currentT1} (b). 
Initially, the current shows a large peak, and then
relaxes to the steady state value $j_0$.
The initial peak is due to vacuum polarization, {\it i.e.}, the electric field 
pulls the quark antiquark pairs apart and the vacuum becomes polarized.
This is better understood in the gapped case ($\eta\ne 0$) 
where the integral of the initial current peak is related to the
induced vacuum polarization via 
$\int_0^\infty j(t)dt=P(E)$, {\it c.f.}, (\ref{eq:PE}). 
However, in the gapless case this integral is ill defined 
and diverges since the current relaxes to a nonzero value $j=j_0$.

Figure \ref{currentT1} (c) shows the time evolution of the 
effective Hawking temperature $T_{\rm eff}$ plotted against the 
boundary time. 
In this example, $T_{\rm eff}$ emerges at the 
thermalization time $t=\tau_{\rm th}$,
and then quickly converges to the 
steady state value $T_{\rm eff}^\infty$.
Although the true steady state is obtained when 
$T_{\rm eff}$ converges to $T_{\rm eff}^\infty$, 
the thermalization time $t=\tau_{\rm th}$
captures the time scale of the whole relaxation process. 
Indeed, the value of the current seems to 
converge to the steady state value $j_0$ already around
$t=\tau_{\rm th}$ as seen in Fig.~\ref{currentT1} (b).

The thermalization time is of particular interest in connection with 
experimental observations in RHIC and LHC that the 
QGP relaxes to the hydrodynamical regime quite fast. 
In Fig.~\ref{thermalizationtime}, we plot $\tau_{\rm th}$ obtained 
for several processes with different parameters (the 
electric fields $E$ and the $AdS$ radius $R$ corresponding to the
coupling constant of the gauge theory).  
In the present gapless case, we expect that the thermalization 
time is related to the steady state effective Hawking 
temperature, which is the most 
prominent energy scale in the long time limit
(the effect of the switch-on timescale $\omega$ 
should disappear by then).
Indeed, by plotting $\tau_{\rm th}$ against $1/z_p$ in Fig.~\ref{thermalizationtime} (b),
we find a relation 
\begin{eqnarray}
\tau_{\rm th}\sim a \frac{\hbar}{ k_B T_{\rm eff}^\infty},
\label{eq:Planckian}
\end{eqnarray}
where $a$ is a non-universal constant around $0.40/\pi$
in our numerical result. 
We have recovered the Planck constant (devided by $2\pi$) $\hbar$ 
and the Boltzmann constant $k_B$.

What is the physical meaning of the effective Hawking temperature
that is induced by the electric field, and 
why does it govern the thermalization time scale?\footnote{
One possibility is the Unruh effect \cite{Unruh1976}.
In a system with charged particles accelerated by the electric fields, 
the Unruh effect  may lead to an induced effective temperature
$T_{\rm Unruh}=\hbar b/2\pi ck_B$, where $b$ is the 
acceleration and  $c$ is the speed of light. 
However, in the present massless case, 
the speed of excitations do not change,
{\it i.e.}, they are fixed to the speed of light.
Thus, it is difficult to relate Unruh effect
to the present system. }
A hint comes from the condensed matter community, 
where similar physics has been studied 
in a zero temperature correlated system at a quantum critical point
\cite{MitraTakeiKimMillis06,Mitra2008}. 
When such systems are driven to a nonequilibrium state with finite current, 
many properties similar to finite temperature systems are observed. 
Using the Keldysh Green's function technique in a model of correlated electrons,it was demonstrated that the 
equations of motion of collective variables
acquire a Langevin dissipation term 
with a Gaussian noise variable $\xi(x,t)$ that 
satisfies $\bra \xi(x,t)\xi(x',t')\ket=\delta(x-x')\delta(t-t')
2T_{\rm eff}^\infty/\gamma$ ($\gamma$ is a model dependent parameter) 
\cite{MitraTakeiKimMillis06}. 
This means that the system behaves as if it was coupled to a 
thermal bath with an effective temperature, in the present case,  
$T_{\rm eff}^\infty$.
This gives a quantum field theory picture for our 
findings in the gravity dual description; 
The effective Hawking temperature is the dual of the 
nonequilibrium-induced Gaussian noise
and acts as a source of decoherence and relaxation,
which was explicitly studied by Sonner and Green \cite{Sonner:2012if}. 
Formula (\ref{eq:Planckian}) relates the 
relaxation time scale to the effective temperature via
two natural unit constants $\hbar$ and $k_B$.
We call this time scale the 
{\it Planckian thermalization time}, which 
becomes shorter (accelerated) as the electric field becomes stronger. 
A similar relation 
was discussed (although it was not proven) in Ref.~\cite{Zaanen2004} 
in the context of equilibrium Hi-Tc superconductors 
where the relaxation time was related to 
the inverse of the transition temperature $T_c$. 
The expression $\hbar/k_B T$ gives the shortest possible relaxation time scale
and is realized when ``Planckian dissipation"
happens, {\it i.e.}, when no other time scale exists 
as in a conformal field theories (quantum critical point)
coupled to a heat bath. 
The present situation is a nonequilibrium version
of the Planckian dissipation where the source of 
dissipation is provided by the induced Gaussian noise
(= fluctuation that is characterized by the effective Hawking temperature)
as explained above.

\FIGURE[hbt]{
\includegraphics[width=14cm]{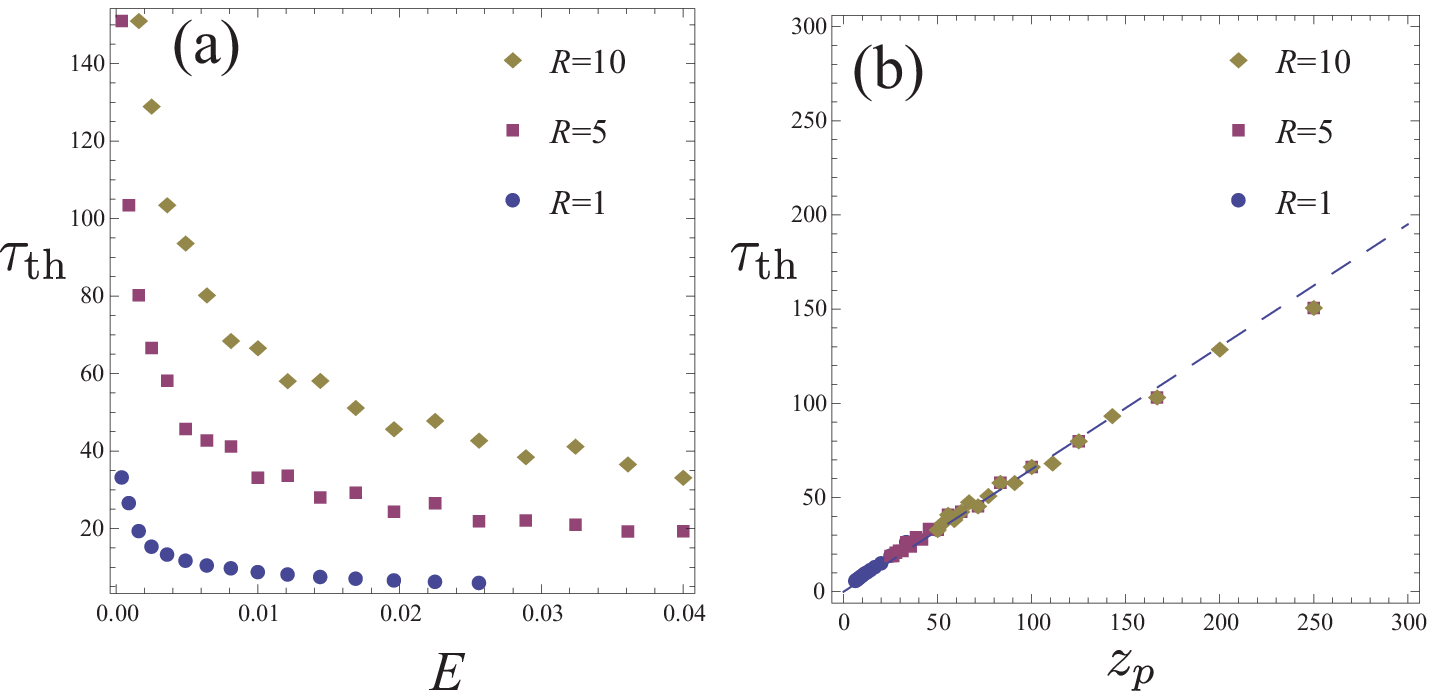}
\caption{
Thermalization time $\tau_{\rm th}$ obtained for 
several processes with 
different field strengths 
and radius $R=1,5,10$. The quench speed is fixed to $\omega=4$.
In (a), it is plotted against the field strength,
and in (b) against the Planckian time scale $z_p$.
An overlap takes place in (b), where
the dashed line denotes the relation (\ref{eq:Planckian}).
}
\label{thermalizationtime}
}

In the present system, the field induced Planckian time
can be very short when the strength of the 
electric field becomes strong. 
This can be relevant in explaining the fast thermalization
problem in RHIC and LHC. 
Of course, in reality, the hadrons are in the 
confined phase of QCD, while our real time calculation
is done in the gapless deconfined phase of a supersymmetric QCD. 
However, when the external color electric field is
stronger than all the other energy scales, 
we expect that (\ref{eq:Planckian}) still sets the thermalization time. 
Let us estimate the induced Planckian time;
Formula (\ref{eq:Planckian}) can be rewritten as ($d=0,\;\eta=0$)
\begin{eqnarray}
\tau_{\rm th}\sim a \pi \left(\frac{\lambda}{2\pi^2}\right)^{1/4}
\frac{\hbar}{ k_B }E^{-1/2}.
\label{eq:Planckian2}
\end{eqnarray}
The typical field strength realized in RHIC is $E=10^4\; [\mbox{MeV}^2]$
\cite{Kharzeev:2007jp,Skokov:2009qp,Voronyuk:2011jd,Bzdak:2011yy,Deng:2012pc}. 
The string tension is an $\mathcal{O}(10)$ 
constant, which we set here to be
$\lambda=10$, 
$\hbar/k_B=4\pi\times 6.08\times 10^{-13}\mbox{Ks}$
($4\pi$ times the universal shear viscosity
\cite{Kovtun2005})
and $300 \mbox{fm/c}\sim 10^{-21}\mbox{s}$.
Thus, we obtain,
\begin{eqnarray}
\tau_{\rm th}\sim
1 \; [\mbox{fm/c}] \, .
\end{eqnarray}
This is of the same order of magnitude, 
compared to the upper bound of the thermalization time 
$1\mbox{fm/s}$ obtained from hydrodynamic analysis of QGP
in RHIC  \cite{Heinz2004}.
However, in spite of this agreement, we must note that 
the present result cannot be the whole story.
The calculation is based on the D3/D7 model in the probe limit.
This means that only the quark dynamics is considered 
and the gluons are always in the thermal
state (zero temperature in this case).
In order to go further, we must consider gluon dynamics
by including backreaction in the gluon sector.

From the viewpoint of dimension analysis, 
(\ref{eq:Planckian2}) is the simplest expression 
to obtain a time scale from the field strength. 
We expect that it gives the universal
lower bound for the thermalization time scale
in QGP production triggered by an external electric field. 
In addition, our finding suggests that 
QGP may reach the hydrodynamic regime in the presence of the field and 
{\it not} after it is lost. 
The hydrodynamic state is a nonequilibrium 
one with finite current, 
and the relaxation to this state may be accelerated by the 
fluctuation characterized by the effective Hawking temperature. 
The duration of the electric field \cite{Deng:2012pc} should be long enougth 
 to observe this effect.

In addition, there are other calculations of the thermalization in AdS/CFT which
attempt to provide an explanation of the early thermalization from a different viewpoint,
see for example Ref.~\cite{Chesler:2008hg} for an early work. 
One needs a more detailed comparison to our calculation.


\section{Conclusion and discussions}

In this paper we studied the vacuum decay process in 
the ${\cal N}=2$  supersymmetric QCD 
at the strong coupling limit
and in the large $N_c$ limit via gauge/gravity duality. 
The decay is induced by strong electric fields 
which the quark hypermultiplet feels. 
We have derived the effective 
nonlinear electromagnetism Lagrangian which is known as
an Euler-Heisenberg Lagrangian in constant background electric field 
to its full order in strong coupling gluons, 
and also derived the imaginary part of the
effective Lagrangian that gives the vacuum decay rate. 
This is possible because the effective Lagrangian
can be expressed as the D-brane's DBI action.

We also calculated the time-dependent response of the system once we apply
a time-dependent electric field. When there is no mass gap, we 
find that the system evolves to a nonequilibrium stationary system with
a constant electric current, in a time-dependent way. The thermalization
time scale is calculated via the gauge/gravity duality, and 
the thermalization time-scale is found to be the Planckian time 
where the temperature  is that of the final stationary state.

It is interesting that the calculated imaginary part of the effective Lagrangian 
for the constant electric field 
coincides with that of the ordinary QED, at the leading and sub-leading order
in the large $E$ expansion. This is a strange coincidence in the sense that
our calculation of course includes all-order gluon effect at the leading 
large 't Hooft coupling, while in QED there is no gluon. 

We finally discuss a relation between our calculations of the imaginary part
and the holographic
Schwinger effect studied in Ref.~\cite{Semenoff2011}. In Ref.~\cite{Semenoff2011}, 
a direct generalization of the Schwinger instanton calculation of 
QED to strongly coupled ${\cal N}=4$ supersymmetric Yang-Mills 
theory was presented. 
There, instead of the electron loop in a Euclidean spacetime 
for QED for the instanton calculation, 
one considers a string worldsheet in the $AdS_5\times S^5$ geometry, 
with a probe D3-brane on which a constant electric field lives. 
The computation is, as in the case of the standard Schwinger calculations, 
valid at weak background electric field 
$E\ll m^2_W$ where $m_W$ is the W-boson mass appearing 
in the Coulomb branch of
the ${\cal N}=4$ supersymmetric Yang-Mills theory. 
The result is, surprisingly, identical
to the original Schwinger's result, with a production rate
\begin{eqnarray}
\Gamma_{\rm open \; string} \sim \exp\left[
-\pi\frac{m_{\rm eff}(E)^2}{ E}
\right] \, ,
\label{eq:SZ}
\end{eqnarray}
where the renormalized  mass is  
\begin{eqnarray}
m_{\rm eff}(E)=m_W-\frac{1}{2}A\frac{E}{m_W}
\end{eqnarray}
with $A=\sqrt{\lambda}/{\pi}$. 
Note that this expression agrees with the 
proposed all loop expression of 
QED effective Lagrangian for the single instanton case
discussed in Refs.~\cite{Ritus78,LebedevRitus84,Affleck1982} (also described in
Ref.~\cite{Dunne2004}).

\FIGURE[r]{ 
\includegraphics[width=7cm]{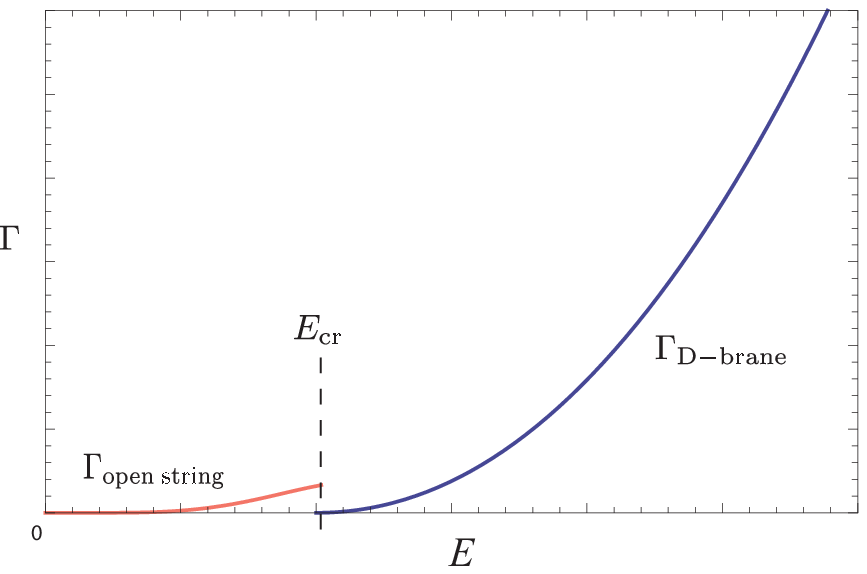}
\caption{Schematic plot of the vacuum decay rate (production rate)
obtained from the open string calculation \cite{Semenoff2011}
and from the D-brane calculation (\ref{eq:EH}).
The true decay rate should be a function which smoothly connects the 
two lines. 
}
\label{semenoff}
}

If we had studied the pair production process in 
a similar manner for our case,
we expect the same result as the
holographic Schwinger effect in Ref.~\cite{Semenoff2011},
 with $m_q = m_W$, because the separated D3-brane in Ref.~\cite{Semenoff2011}
is just replaced by our flavor D7-brane.
Thus, the expression (\ref{eq:SZ}) is  also 
valid in our ${\cal N}=2$ supersymmetric QCD.
If we plot the production
rate in the string and D-brane sectors
we find a contradictory-looking fact; the worldsheet instanton
(\ref{eq:SZ}) gives a nonzero production rate for any small $E$, 
while our calculation using the D-brane action 
(\ref{eq:EH}) has a critical electric field 
below which the production rate vanishes.
The resolution of this ``puzzle" is as follows: 
the true decay rate should be 
given by a function which smoothly connects the two solutions.
This is because 
both calculations actually evaluate the same disk partition 
function, but in different limits.
The holographic Schwinger effect (\ref{eq:SZ}) based on the disk 
amplitude evaluated in the $AdS_5$ is valid only at weak electric field. 
When electric field is weak,
the area of the worldsheet instanton is large, and cannot be captured
by its flat-space analogue at which any disk would be 
stabilized to a flat small one. 
On the other hand, the Dirac-Born-Infeld action which we used in this paper
is based on a disk amplitude with a flat and small disk worldsheet. 
In the $AdS_5$ geometry, the small disk is realized in the large electric
field limit as worldsheet instantons, so our DBI analysis is valid 
at large $E$. 
So, the two expressions come from the same single disk partition
 function with $E$ vertex operator insertions, but in different limits.

To gain more insight on the interplay between
the two limits, in Fig.~\ref{semenoff} we plot 
the two contributions in a single plot.
For small $E$, the string result (\ref{eq:SZ}) gives a nonzero but
exponentially suppressed production rate, while 
 at large $E$, we have a power-law large imaginary part. 
We expect that the whole behavior would be a smooth combination of
the two lines. 

We have seen in section 4 that our result of the imaginary part of the
Lagrangian reproduces the leading and sub-leading behavior of the QED
Schwinger effect calculations. How can it be consistent with (\ref{eq:SZ})?
In fact, the agreement which we found in section 4 is not really (\ref{eq:SZ})
but a different exponent, $m^2/(\sqrt{\lambda}E)$. See eq.~(\ref{eq:imexpm})
where one finds $E_{\rm cr} \sim m^2/\sqrt{\lambda}$, instead of $m^2$
in (\ref{eq:SZ}). Since the gauge/gravity duality is valid at the large
'tHooft coupling constant $\lambda$, the holographic Schwinger effect
(\ref{eq:SZ}) has a much more suppression than what we found in section
4, by the factor $\sqrt{\lambda}$ in the exponent.

There is another lesson which we can learn from the holographic Schwinger effect
(\ref{eq:SZ}). A natural question  arising in our derivation of the imaginary part is:
why could we get the imaginary part by a classical action (string disk amplitude) 
rather than quantum loops? In fact, in the evaluation in a flat target spacetime in string
theory, the imaginary part came from a cylinder amplitude which is a quantum loop
\cite{Burgess:1986dw,Nesterenko:1989pz,Bachas1992a,Bachas:1995kx}. 
We can find a resolution of this counter-intuitive fact, by looking
at the holographic Schwinger effect \cite{Semenoff2011}. In the gauge/gravity duality,
one replaces the $N_c$ D3-branes by the $AdS_5$ geometry. Removing the D3-branes means that one needs to cap off one boundary of the cylinder, so
it results in a disk worldsheet, that is, a classical DBI action. So, we learn 
that the reason why the classical DBI action can give the imaginary part is
the gauge/gravity duality. 

Finally, it is obvious 
that we have many future directions to extend our way to calculate the instability and the full Euler-Heisenberg action
for strongly coupled systems. We would like to report on further applications
elsewhere.


\acknowledgments
K.H.~is indebted to Hirosi Ooguri for valuable discussions and comments
which were indispensable for finishing this work.
K.H.~would also 
like to thank K.~Fukushima, G.~Semenoff, H.~Suganuma and S.~Sugimoto
for discussions. K.H.~ is grateful to the hospitality of Caltech theory group, and 
APCTP.
T.O~ acknowledeges Y.~Hidaka and S.~Sasaki for insightful discussions. 
K.H.~and T.O.~are partly supported by
the Japan Ministry of Education, Culture, Sports, Science and
Technology. 
This research was partially supported by the RIKEN iTHES project.

\appendix

\section{The imaginary action at zero temperature}
\label{app:T0}

Our result (\ref{eq:imld0}) for the gapless system is a bit complicated,
so here we perform some more calculations for a simpler $T=0$ case.
We set $z_{\rm H}=\infty$, then (\ref{eq:imld0}) becomes 
\begin{eqnarray}
{\rm Im} {\cal L} = 2\pi^2 \mu_7 \int_{R/\sqrt{2\pi\alpha' E}}^\infty dz
\frac{R^8}{z^5}\sqrt{\left(\frac{(2\pi\alpha' E)^2}{R^4} z^4-1\right)
\left(1 + \frac{z^6}{R^4}d^2\right)^{-1}}\, .
\end{eqnarray}
To see the dependence in physical quantities, 
we change the integration coordinate as
\begin{eqnarray}
z^2 = \frac{R^2}{2\pi\alpha' E}\frac{1}{x}\, ,
\end{eqnarray}
to obtain
\begin{eqnarray}
{\rm Im} {\cal L} = \pi^2 \mu_7 \frac{(2\pi\alpha' E)^2}{R^4}
\int_0^1 dx \; x^{3/2}\sqrt{\frac{1-x^2}{x^3 + d^2R^2/(2\pi\alpha' E)^3}} \, .
\end{eqnarray}
{}From this expression, we can see that, for a given electric field, a larger charge 
density means a smaller magnitude of the imaginary part.
This is counter-intuitive. In fact, when the imaginary part is zero,
the electric current is given by (\ref{eq:jT0}) which shows a 
larger current for a larger density $d$. So, the $d$-dependence of 
the instability does not directly relate to the $d$-dependence of the
electric current $j$. 

To see this fact in more detail, we perform a small $d$ expansion
of the two quantities: Im ${\cal L}$ and the current $j$. First, we have
\begin{eqnarray}
{\rm Im} {\cal L} = \frac{\pi}{4} \mu_7 \frac{(2\pi\alpha' E)^2}{R^4}
\left[
1-\frac{4c}{\pi} d^{2/3} \frac{R^{2/3}}{2\pi\alpha' E}
+ \mbox{higher order in $d$}\right] \, ,
\end{eqnarray}
with a numerical coefficient $c \simeq 0.862$. On the other hand, the current $j$ can be expanded for small density $d$ as
\begin{eqnarray}
j = \frac{(2\pi\alpha' E)^{3/2}}{R}
\left[
1+ \frac{1}{2}\frac{R^2}{(2\pi\alpha' E)^3} d^2 
+ \mbox{higher order in $d$}\right] \, .
\end{eqnarray}
The dependence for small $d$ is totally different.

Normally, a larger instability would lead to a larger current as a consequence of
the vacuum breakdown. However, here in our supersymmetric QCD at strong coupling, the situation is not that intuitive.

\section{The induced metric and the volume}
\label{app:volume}

In this appendix, we present the calculation of the induced metric 
on the D7-brane under the time-dependent electric field, studied in
Sec.~\ref{sec:td}. We derive eq.~(\ref{eq:volume}) and eq.~(\ref{eq:null}).
Furthermore, we shall derive the effective 
Hawking temperature (\ref{eq:effHaw}) by 
using the induced metric. (See for example
Refs.~\cite{Kim:2011qh,Sonner:2012if} for a similar calculation for a D3D5 steady system.)

The induced metric in the absence of the electric field on the D7-brane is
\begin{eqnarray}
P[g]_{\mu\nu}= \frac{R^2}{z^2} \eta_{\mu\nu} \, , \quad
P[g]_{zz} = \frac{R^2}{z^2} \, , \quad P[g]_{IJ}=R^2 G_{IJ}^{S^3}\, .
\end{eqnarray}
The D7-brane action includes $\det(P[g] + 2\pi\alpha' F)$, and we expand it
around a time-dependent electric field background.
Defining the background $F_{ab}$ plus the original induced metric $P[g]_{ab}$ 
as $M_{ab}$, 
\begin{eqnarray}
M_{ab} \equiv P[g]_{ab} + 2\pi\alpha' F^{\rm sol}_{ab} \, ,
\end{eqnarray}
and denote the fluctuation field strength as just $\delta F_{ab}$, then
\begin{eqnarray}
&&\lefteqn{
\sqrt{-\det (M+2\pi\alpha' \delta F)}
}
\nonumber \\
&&= \sqrt{-\det M} 
\left[
1 + \frac{(2\pi\alpha')^2}{8}\left(
\left({\rm tr}[M^{-1} \delta F]\right)^2
-2 {\rm tr}[M^{-1} \delta F M^{-1} \delta F]
\right)
+ {\cal O}((\delta F)^3)
\right] \, .
\nonumber
\end{eqnarray}
Here, we kept $\delta F$ to its second order, and the first order fluctuation
vanishes due to the fact that we are expanding around a classical solution.

Now, note that $M^{-1}$ includes an anti-symmetric part as well as a symmetric part. We decompose it as 
$M^{-1} \equiv (M^{-1})^s + (M^{-1})^a$, where ``s" and ``a" mean symmetric and antisymmetric components, 
respectively.  Then, assuming that the background solution
depends only on $t$ and $z$ and also that the background field strength is nonzero 
only with the components $F_{01}$, $F_{0z}$ and $F_{1z}$, we can simplify the fluctuation algebraically as
\begin{eqnarray}
\sqrt{-\det (M+2\pi\alpha' \delta F)}
\biggm|_{{\cal O}((\delta F)^2)}
= -\sqrt{-\det M} 
\frac{(2\pi\alpha')^2}{4}
 {\rm tr}[(M^{-1})^s \delta F (M^{-1})^s \delta F]
 \, .
\nonumber
\end{eqnarray}
The front factor can be decomposed as
\begin{eqnarray}
\sqrt{-\det M} 
= \sqrt{-\det ((M^{-1})^s)^{-1}} \times \gamma\, , \quad
\gamma \equiv \sqrt{\det (1 + ((M^{-1})^s)^{-1} (M^{-1})^a)} \, .
\end{eqnarray}
So, writing
\begin{eqnarray}
G_{ab} &\equiv&
\mu_7^{1/2} \gamma^{1/2} (2\pi\alpha')
 \left[\left((M^{-1})^s\right)^{-1}\right]_{ab}\, ,
 \label{eq:effG}
\end{eqnarray}
we obtain the fluctuation Lagrangian as
\begin{eqnarray}
-\mu_7 \sqrt{-\det (M+2\pi\alpha' \delta F)}
\biggm|_{{\cal O}((\delta F)^2)}
= \frac{1}{4}\sqrt{-\det G}\,  
 {\rm tr}[G^{-1}\delta F G^{-1} \delta F]
 \, .
\end{eqnarray}
This is nothing but the standard Maxwell Lagrangian, so this $G$ given by
eq.~(\ref{eq:effG}) can be regarded 
as a metric which the gauge fluctuation feels.

Once the effective metric $G$ is derived, the volume element (\ref{eq:volume}) 
which is necessary
to calculate the location of the apparent horizon can be easily obtained. 
Here we present the explicit form of the effective metric $G_{ab}$. 
First, the $(0,1,z)$ sector of the metric is
\begin{eqnarray}
G = \frac{R^2}{z^2}
\left(
\begin{array}{ccc}
-1+a^2+b^2 & bc & -ac \\
bc & 1-a^2+c^2 & -ab \\
-ac & -ab & 1-b^2 + c^2
\end{array}
\right)
\label{eq:effG}
\end{eqnarray}
where
\begin{eqnarray}
a \equiv 2\pi\alpha' \frac{z^2}{R^2} F_{01} \, , 
\quad
b \equiv 2\pi\alpha' \frac{z^2}{R^2} F_{0z} \, , 
\quad
c \equiv 2\pi\alpha' \frac{z^2}{R^2} F_{1z} \, .
\end{eqnarray}
The $(2,3)$ sector is
\begin{eqnarray}
G = \frac{R^2}{z^2}
\left(
\begin{array}{cc}
1 & 0 \\
0 & 1
\end{array}
\right) \, ,
\end{eqnarray}
and the $S^3$ sector is given by $G_{IJ} = R^2 G_{IJ}^{\rm unit \; S^3}$.

{}From this formula for the induced metric, let us obtain an explicit expression 
of the metric for the nonequilibrium steady state with the electric current 
considered in section 2. This can derive the effective Hawking temperature (\ref{eq:effHaw}). 
For simplicity, we take $d=0$ (vanishing charge density) and $T=0$ 
(zero temperature for the bulk gluons). 
The configuration of section 2 provides
\begin{eqnarray}
 a = \frac{z^2}{z_p^2} \, , \quad b=0 \, , \quad 
 c = \frac{R^2 z}{z_p^3} \sqrt{\frac{1-z^4/z_p^4}{1-z^6/z_p^6}} \, .
 \label{eq:abc}
\end{eqnarray}
The nontrivial $(0,z)$ components of the effective metric $G$ (\ref{eq:effG})
is given as
\begin{eqnarray}
ds^2 = \frac{R^2}{z^2}
\left(
(-1+a^2) dt^2 -2ac \; dt \; dz + (1+c^2)dz^2
\right)\, .
\end{eqnarray}
We can diagonalize this metric by the following field redefinition 
(see for example a similar transformation in Ref.~\cite{Sonner:2012if})
\begin{eqnarray}
t' \equiv t + \int^z \!\! dz \frac{ac}{1-z^2} 
\end{eqnarray}
to obtain
\begin{eqnarray}
ds^2 = \frac{R^2}{z^2}
\left(
(-1+a^2) dt'^2 + \frac{-1+a^2-c^2}{-1+a^2}dz^2
\right)\, .
\end{eqnarray}
Substituting (\ref{eq:abc}) and expanding the metric around 
$z_p$ as $z =z_p + \delta z$ (here $\delta z < 0$), 
\begin{eqnarray}
ds^2 = \frac{R^2}{z_p^2}
\left(-
\frac{2|\delta z|}{z_p}
dt'^2 +\frac{z_p}{|\delta z|}[d(\delta z)]^2
\right)\, .
\end{eqnarray} 
This metric is nothing but a near-horizon metric of a black hole.
So we can actually see that the mesons on the D7-brane feels a metric
with an event horizon.

Using a redefinition to a radial $r$ coordinate,  $|\delta z| \equiv (3z_p/4R^2) r^2$,
and Euclideanize the time direction $t'$ and imposing the smoothness condition
of the metric, we obtain the periodicity
\begin{eqnarray}
i t' \sim i t' + 2\pi z_p \sqrt{2/3} \, . 
\end{eqnarray}
So the Hawking temperature calculated from the induced metric is given by
\begin{eqnarray}
T_{\rm eff} = \sqrt{\frac{3}{8}} \frac{1}{\pi z_p} \, .
\end{eqnarray}
This is the effective temperature which the mesons feel, in the stationary 
nonequilibrium case, (\ref{eq:effHaw}).
In section 5, we have defined a time-dependent version of the effective temperature (\ref{eq:Teff}),
by just generalizing (\ref{eq:effHaw}). The definition of (\ref{eq:Teff}) uses the
apparent horizon, so it is a natural choice, among ones which converges to
(\ref{eq:effHaw}) at later time.

\bibliographystyle{JHEP.bst}
\bibliography{hep-th130724.bib}

\end{document}